\documentclass[fleqn,usenatbib]{mnras}

\usepackage[T1]{fontenc}
\usepackage{ae,aecompl}

\usepackage{graphicx}	
\usepackage{amsmath}	
\usepackage{amssymb}	
\usepackage{tikz}
\usepackage{xcolor}

\usepackage{fontawesome}
\usetikzlibrary{arrows,shapes,positioning}

\DeclareMathOperator*{\argmax}{arg\,max}
\usepackage{esvect} 

\usepackage{xcolor}

\newcommand{\github}{\href{https://github.com/NiallJeffrey/Likelihood-free_DES_SV}{\faGithub}}

\title[Likelihood-free DES SV map statistics]{Likelihood-free inference with neural compression of DES SV weak lensing map statistics}

\author[N. Jeffrey et al.]{Niall Jeffrey$^{1,2}$\thanks{E-mail: \href{niall.jeffrey@phys.ens.fr}{niall.jeffrey@phys.ens.fr}}, Justin Alsing$^{3}$,  
Fran\c{c}ois~Lanusse$^{4}$
\\
$^{1}$Laboratoire de Physique de l'Ecole Normale Sup\'erieure, ENS, Universit\'e PSL, CNRS, Sorbonne Universit\'e, Universit\'e de Paris,\\
Paris, France \\
$^{2}$Department of Physics \& Astronomy, University College London, Gower Street, London, WC1E 6BT, UK\\
$^{3}$Oskar Klein Centre for Cosmoparticle Physics, Department of Physics, Stockholm University, Stockholm SE-106 91, Sweden \\
$^{4}$AIM, CEA, CNRS, Universit\'e Paris-Saclay, Universit\'e Paris Diderot, Sorbonne Paris Cit\'e, F-91191 Gif-sur-Yvette, France
}

\date{Accepted 2020 November 12. Received 2020 October 26; in original form 2020 August 18}

\pubyear{2020}
\begin{document}
\label{firstpage}
\pagerange{\pageref{firstpage}--\pageref{lastpage}}
\maketitle

\begin{abstract}
In many cosmological inference problems, the likelihood (the probability of the observed data as a function of the unknown parameters) is unknown or intractable. This necessitates approximations and assumptions, which can lead to incorrect inference of cosmological parameters, including the nature of dark matter and dark energy, or create artificial model tensions. Likelihood-free inference covers a novel family of methods to rigorously estimate posterior distributions of parameters using forward modelling of mock data. We present likelihood-free cosmological parameter inference using weak lensing maps from the Dark Energy Survey (DES) SV data, using neural data compression of weak lensing map summary statistics. We explore combinations of the power spectra, peak counts, and neural compressed summaries of the lensing mass map using deep convolution neural networks. We demonstrate methods to validate the inference process, for both the data modelling and the probability density estimation steps. Likelihood-free inference provides a robust and scalable alternative for rigorous large-scale cosmological inference with galaxy survey data (for DES, Euclid and LSST). { We have made our simulated lensing maps publicly available.} \github
\end{abstract}




\section{Introduction}
Likelihood-free inference allows us to infer unknown cosmological parameters by directly comparing observed data with forward-simulated mock data. In this powerful and flexible framework, the posterior probability of unknown parameters can be estimated without resorting to simplifying, and often unjustified, likelihood assumptions (such as Gaussianity), even when data models consist of complex combinations of signal, noise, and systematic error. Incorrect assumptions can lead to incorrect and misleading conclusions for the given cosmological question, can hide new physics, or can create artificial model tensions.

The aim of observational cosmology is often to infer the cosmological parameters or models from the structure of the density field of the Universe as it evolves with time. For the late Universe, non-linear evolution has led to a highly non-Gaussian density field, which cannot be statistically characterized solely by 2-point statistics (e.g. power spectra). The gravitational lensing effect on images of distant galaxies by the intervening large-scale structure provides a powerful probe of cosmology in this regime, through both structure formation and the geometry of the Universe.

In this work we use measured statistics of the reconstructed projected density field (known as \textit{mass maps}) from Dark Energy Survey (DES,~\citealt{decam},~\citealt{more_than_de}) weak gravitational lensing data to infer cosmological parameters of the $\Lambda$CDM model in a likelihood-free analysis. We use deep learning methods with the aim of extracting the optimal compressed statistic from our chosen data/statistic; this method is known as \textit{neural compression}. 

The physics of non-linear cosmological structure formation is included in our forward-modelled mock data using simulated (approximate $N$-Body) density fields, with the lensing map observables calculated through ray tracing, and subsequent inclusion of (complicated) masks and non-Gaussian noise contributions corresponding to the DES data. The posterior probability densities for unknown parameters $\boldsymbol{\theta}$ are then estimated for different lensing map statistics from observed (unsimulated) data $\mathbf{d}_o $, without the need for an assumed analytic expression for the likelihood function $\mathcal{L}(\boldsymbol{\theta}) = p(\mathbf{d}_o | \boldsymbol{\theta})$. 

The map-based statistics used in this likelihood-free analysis are: (1) the angular power spectrum of the map;   (2) the peak statistics, also known as peak counts (\citealt{dietrich_peaks};~\citealt{des_sv_peaks});   \noindent (3) the joint statistic of peaks and power spectrum;   \noindent  (4) convolutional neural network (CNN) compressed map statistics, aiming to compress the reconstructed mass map to optimal summary statistics using deep learning (\citealt{fluri_cnn};~\citealt{ribli_cnn};~\citealt{kids_cnn}).

We note that the choice of mapping method, in our case Kaiser-Squires~\citep{ks93}, effectively corresponds to an initial data compression step, compressing a catalogue of images into a pixelized estimated mass map. Note that peak counts estimated from a different map reconstruction method may lead to somewhat different constraints on the cosmological parameters, owing to different information retainment associated with different reconstructions methods. 

Likelihood-free inference provides an alternative inference framework for current and upcoming galaxy surveys (e.g. Euclid:~\citealt{euclid_science}, or the Vera C. Rubin Observatory's Legacy Survey of Space and Time (LSST):~\citealt{lsst_science}), relying only on our ability to forward-simulate mock data and unfettered by the need for a closed form likelihood function. Given their appeal, it is important to understand the challenges presented by such methods, and to establish procedures to carefully validate the results. 

Given current reported tensions in cosmological parameters between different data sets, such as the Hubble parameter $H_0$ between early- and late-Universe probes~\citep{hubble_tension} or the amplitude of fluctuations between galaxy surveys (\citealt{des_year1};~\citealt{kids18}) and the Cosmic Microwave Background~\citep{planck18}, likelihood-free inference offers a novel inference framework without restrictive assumptions about the likelihood or data-model. Likelihood-free inference also provides a simple and robust framework for combining observed summary statistics from different data sets with forward modelling, and avoids many of the potentially difficult technical aspects of current standard analysis (such as covariance matrix estimation or sampling from high-dimensional Bayesian hierarchical models).

\begin{figure}
\hspace*{-0.5cm}
\vspace*{-0.5cm}
\includegraphics[width=0.513\textwidth]{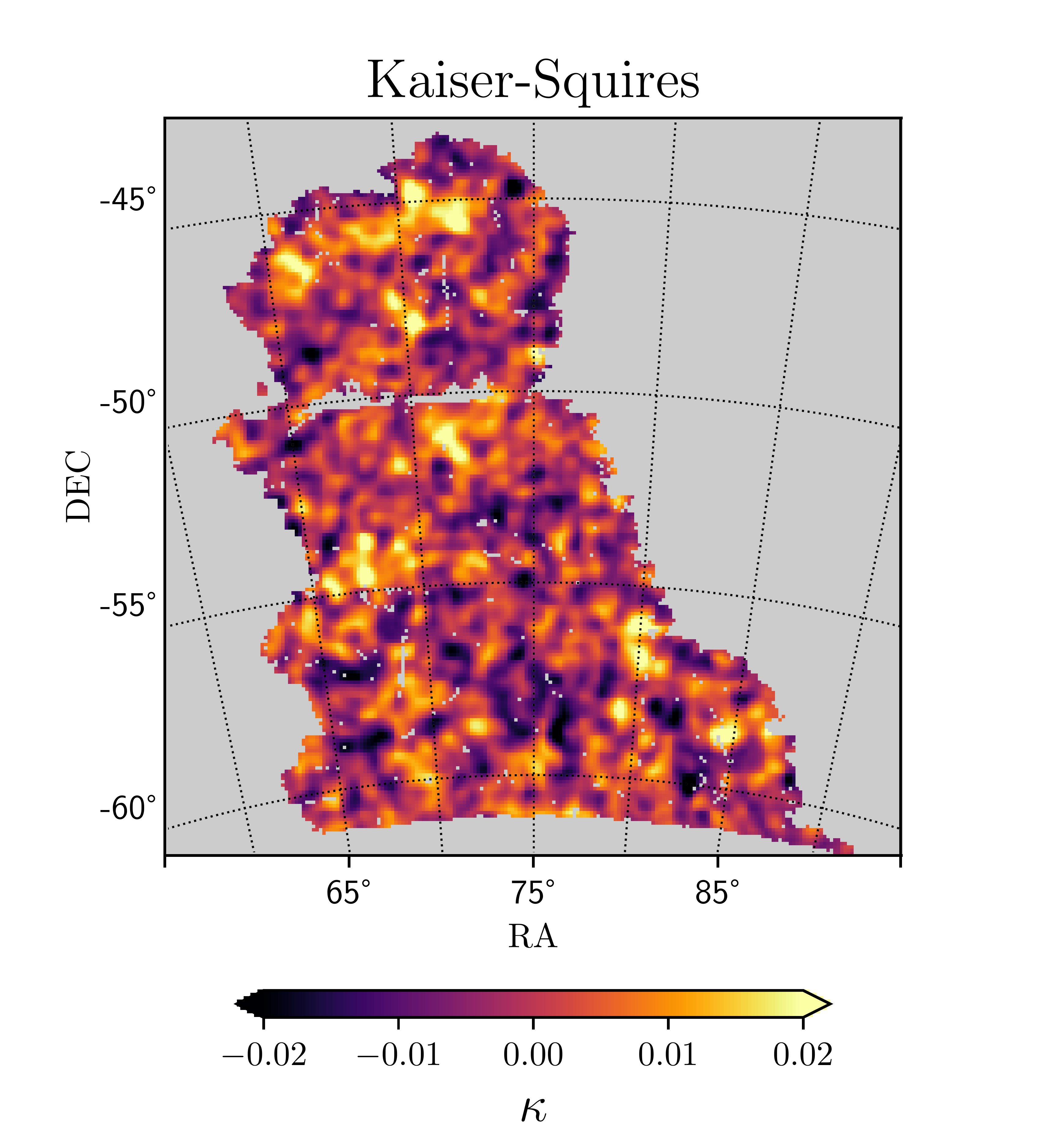}
\caption{\label{fig:ks} Convergence $\kappa$ map from the DES SV weak lensing data reconstructed using Kaiser-Squires with $\sigma=10\ {\rm{arcmin}}$ Gaussian smoothing to reduce the impact of galaxy shape noise.}
\vspace*{0.1cm}
\end{figure}

The mass maps in this work (Figure~\ref{fig:ks}) are generated from public DES Science Verification (SV) data~\citep{chang_sv_map} using the linear Kaiser-Squires method. Though the SV data cover an area of only approximately 5 per cent of the final DES sky footprint, the observations are to the approximate final depth of the full survey, so the SV data match what will be the final galaxy density and lensing signal-to-noise per pixel. 

In Section~\ref{sec:lfi}, we introduce the formalism of likelihood-free inference and data compression methods. In Section~\ref{sec:lensing}, we include an overview of the relevant aspects of weak gravitational lensing with galaxies. In Section~\ref{sec:statistics} we detail our map reconstruction method and the three summary statistics used: power spectrum, peak statistics, and CNN compressed map summaries. In Section~\ref{sec:simulations} we discuss the DES SV data and the forward modelling of mock simulated data. In Section~\ref{sec:results} we discuss the results and validation of the likelihood-free parameter inference using power spectra and peak statistics. In Section~\ref{sec:cnn_result} we discuss results using likelihood-free inference with compressed summaries directly from the mass map using convolutional neural networks.
\section{Inference} \label{sec:lfi}
\subsection{Motivation}
In Bayesian parameter inference we aim to evaluate the posterior probability distribution

\begin{equation} \label{eq:bayes}
    p(\boldsymbol{\theta} | \mathbf{d}_o , \mathcal{M}) = \frac{p(\mathbf{d}_o | \boldsymbol{\theta}, \mathcal{M}) \  p(\boldsymbol{\theta} | \mathcal{M})}{ p(\mathbf{d}_o | \mathcal{M})}
\end{equation}

\noindent for some statistical model $\mathcal{M}$ with associated unknown model parameters $\boldsymbol{\theta}$, given some observed data (or summary statistics of the observed data) $\mathbf{d}_o$ (see~\citealt{jaynes07} for details).

To evaluate the relative probabilities of different models given the observed data summaries, one may also wish to evaluate the Bayesian evidence, also known as the marginal likelihood, given by

\begin{equation}
    p(\mathbf{d}_o | \mathcal{M}) = \int p(\mathbf{d}_o | \boldsymbol{\theta}, \mathcal{M}) \  p(\boldsymbol{\theta} | \mathcal{M}) \ \mathrm{d}^n \boldsymbol{\theta} \ \ 
\end{equation}

\noindent for probability densities of continuous variables $\boldsymbol{\theta}$.

For both of these tasks, knowledge of the likelihood function $\mathcal{L} (\boldsymbol{\theta}) = p(\mathbf{d}_o | \boldsymbol{\theta}, \mathcal{M})$ -- the sampling distribution for the observations as a function of the model parameters -- is required. In this paper we will focus on parameter inference. Despite the central role of the likelihood function, in general the sampling distribution $p(\mathbf{d} | \boldsymbol{\theta})$ for data (or summary statistics) $\mathbf{d}$ is not necessarily readily available (or tractable).

For parameter inference from large cosmological surveys, especially weak lensing surveys (\citealt{cfhtlens}; \citealt{kids18}; \citealt{des_year1}), the conditional distribution $p(\mathbf{d} | \boldsymbol{\theta}, \mathcal{M})$ is not generally known exactly, owing to non-linear evolution of the underlying density field, and any number of complicated observational effects (survey masks, various systematic biases, non-Gaussian noise contributions, etc.). For two-point statistics of the weak lensing field (see Section~\ref{sec:lensing}), a Gaussian sampling distribution is typically assumed\footnote{Exceptions are Bayesian hierarchical analyses, which account for the non-Gaussian data model \citep{alsing2016hierarchical, alsing2017cosmological}.}, though the second-order moments (i.e. two-point statistics) have a skewed distribution even for an underlying Gaussian lensing field (\citealt{sellentin_non_Gaussian};~\citealt{sellentin_skewed};~\citealt{taylor_delfi_lensing}).

For higher-order statistics of the lensing field, which are necessary to extract information beyond the Gaussian component of the field, there is typically no closed-form expression for their sampling distribution (and hence likelihood function), inhibiting their use for cosmological parameter inference. Even theoretical predictions for the expectation values of higher-order statistics (e.g. for peak counts or deep CNN map summary statistics) must be estimated by forward-simulated mock realizations of the data. Sampling distributions for higher-order statistics are not expected to be Gaussian, with non-Gaussianity arising from non-linear combinations (e.g. counting peaks or deep CNN statistics) of an already non-Gaussian cosmological lensing field, compounded by non-Gaussian shape-noise and Poissonian shot-noise in the data.

Arguments that rely on the central-limit theorem or the principle of maximum entropy, both of which give the Gaussian distribution special status,\footnote{See~\cite{jaynes07} for details.} cannot avoid the fact that if one assumes an incorrect distribution $p(\mathbf{d} | \boldsymbol{\theta}, \mathcal{M})$, the resulting statistical inference may be misleading or biased.

\subsection{Likelihood-free inference with forward modelling}

Contrary to what the name might suggest, likelihood-free inference (also known as simulation-based inference) does not exclude a likelihood. The distribution $p(\mathbf{d} | \boldsymbol{\theta})$\footnote{From here on, we will drop the explicit model $\mathcal{M}$ dependence for brevity.} is not used in closed-form, but is reconstructed from simulated mock data as part of the inference pipeline.

In density estimation likelihood-free inference (\citealt{Papamakarios_lfi};~\citealt{delfi1}), the inference task is posed as a density estimation problem. One can picture the simulated mock data $\mathbf{d}$ realizations and their associated parameters $\boldsymbol{\theta}$ as forming a cloud of points in $\{ \mathbf{d}, \ \boldsymbol{\theta} \}$ space. In this space we could estimate the following distributions: (A) the joint $p( \mathbf{d}, \boldsymbol{\theta})$; (B) the conditional $p(  \boldsymbol{\theta} | \mathbf{d})$, which would give the posterior if evaluated for observed data $\mathbf{d}_o$; (C) the conditional $p( \mathbf{d} | \boldsymbol{\theta})$, which would give the likelihood if evaluated for observed data $\mathbf{d}_o$. All of the above can be straightforwardly used to reconstruct the desired posterior density.

Estimating either (A) $p( \mathbf{d}, \boldsymbol{\theta})$ or (B) $p(  \boldsymbol{\theta} | \mathbf{d})$, which are both densities with respect to $\boldsymbol{\theta}$, requires that the distribution of $\boldsymbol{\theta}$ in the set of simulated mocks must come from the prior distribution $p(  \boldsymbol{\theta})$ (see ~\citealp{delfi2}). To avoid this constraint, one can take the strategy (C) of estimating the sampling distribution $p( \mathbf{d} | \boldsymbol{\theta})$ as a function of the model parameters, which is a probability density in $\mathbf{d}$ rather than $\boldsymbol{\theta}$. This has the advantage that the simulated parameter values do not need to be drawn from the prior, enabling the use of various strategies for optimizing (and reducing) the set of simulations to be run for the problem at hand. It also allows for seamless subsequent analyses of alternative prior assumptions once the sampling distribution (and hence likelihood) has been learned.

Once the density $p( \mathbf{d} | \boldsymbol{\theta})$ is learned from our simulated mocks, it can be evaluated at the observed data $\mathbf{d}_o$ and treated as a likelihood $\mathcal{L}(\boldsymbol{\theta})$ in the usual way for parameter inference (equation~\ref{eq:bayes}). To the extent that the physics and the effects of the realistic data model are correctly included in the forward simulations, the estimated $p(\mathbf{d} | \boldsymbol{\theta})$ will be the correct distribution to be used for parameter inference (provided sufficiently many simulations to accurately learn the sampling distribution).

{
The density estimation approach to likelihood-free inference is an alternative to the more traditional Approximate Bayesian Computation (ABC,~\citealt{rubin1984bayesianly};~\citealt{ABC2}), which uses (adaptive) rejection-based sampling (for recent applications in astronomy, see \citealp{Schafer2012,Cameron2012,Weyant2013,Robin2014,Lin2015,Ishida2015, Akeret2015,Davies2016, Jennings2016,Hahn2017,Kacprzak2017,Carassou2017,tortorelli_abc, fagioli_abc}).

In ABC, one draws parameters from some (typically adaptive) proposal density, and forward simulates data. Those simulated data are then compared to the observed data under some distance metric, and the proposed parameters accepted if the distance is below some threshold $\epsilon$. The resulting accepted samples then constitute samples from an approximate posterior, approaching the exact posterior only in the (unattainable limit) $\epsilon\rightarrow0$. In practice, even with sophisticated adaptive methods the use of rejection sampling means that the vast majority of samples get rejected, making for a simple but simulation inefficient approach. 

Density-estimation likelihood-free methods overcomes this limitation by using all simulated data for improving the inference, and does not require a (subjective) distance metric and threshold, allowing for high-fidelity posterior inference with far fewer simulations \citep{Papamakarios_lfi, delfi1}.
}

\subsection{Neural density estimation}

Here we introduce the main mathematical aspects of the neural density estimators used in this work. For a more comprehensive discussion see~\cite{delfi2}. Alternatively, the reader can proceed to Section~\ref{sec:compression_intro} to skip the technical details of neural density estimation.

To estimate the conditional distribution $p( \mathbf{d} | \boldsymbol{\theta})$, we use the pyDELFI~\citep{delfi2} package (see Appendix~\ref{append:codes}) with an ensemble of neural density estimators (NDEs): neural network parameterizations of conditional probability densities. Specifically we use a combination of Gaussian Mixture Density Networks (MDN;~\citealt{mdn}) and Masked Autoregressive Flows (MAF;~\citealt{MAF}). 

{
These are, of course, not the only choice of neural density estimator. For example, \cite{rivero_ffjord} recently demonstrated the characterization of distributions of weak lensing power spectra using the alternative \textsc{ffjord}\footnote{Free-form Continuous Dynamics for Scalable Reversible Generative Models} neural density estimator \citep{ffjord} (among other applications). The vibrant field of neural density estimation, Normalizing Flows in particular, will likely lead to further breakthroughs for probability density estimation and likelihood-free inference in the near future~\citep{papamakarios2019normalizing}.

For both of our neural density estimation methods, MDN and MAF,} the networks are trained to give an estimate $q  (\mathbf{d} |  \boldsymbol{\theta} ; \boldsymbol{\varphi})$ of the target distribution $p( \mathbf{d} | \boldsymbol{\theta})$, interpretable as 

\begin{equation}
    p( \mathbf{d} | \boldsymbol{\theta}) \approx q  (\mathbf{d} |  \boldsymbol{\theta} ;  \boldsymbol{\varphi}) \ \ ,
\end{equation}

\noindent by varying the $\boldsymbol{\varphi}$ parameters (e.g. weights and biases) of the network.\footnote{See~\citealt{deep_learning} for an introduction to neural networks.} This is achieved by minimizing the loss function

\begin{equation} \label{eq:delfi_loss}
U(\boldsymbol{\varphi}) = - \sum_{n=1}^N \log q  (\mathbf{d}_n |  \boldsymbol{\theta}_n ; \boldsymbol{\varphi}) \ \ 
\end{equation}

\noindent over the $N$ forward-modelled mock data $\mathbf{d}_n$. This loss corresponds to minimizing the Kullback-Leibler divergence~\citep{kullback1951}, a measure of difference or change going from the estimate $q$ to the target $p( \mathbf{d} | \boldsymbol{\theta})$.

Gaussian mixture density networks (MDNs) represent the conditional density as a sum of $K$ Gaussian components with mean $\boldsymbol{\mu}(\boldsymbol{\theta}; \boldsymbol{\varphi})_k$, covariance $\mathbf{C}(\boldsymbol{\theta}; \boldsymbol{\varphi})_k$, and component weights $\mathbf{r}(\boldsymbol{\theta}; \boldsymbol{\varphi})_k$ all taken as unknown functions of the parameters $\boldsymbol{\theta}$, parameterized by a neural network:

\begin{equation}
    q  (\mathbf{d} |  \boldsymbol{\theta} ;  \boldsymbol{\varphi}) = \sum_{k=1}^{K} \mathbf{r}(\boldsymbol{\theta}; \boldsymbol{\varphi})_k \mathcal{N} \big[\mathbf{d} \  | \ \boldsymbol{\mu}(\boldsymbol{\theta}; \boldsymbol{\varphi})_k ,\   \mathbf{C}(\boldsymbol{\theta}; \boldsymbol{\varphi})_k \big] \ \ .
\end{equation}

The second density estimation method uses Normalizing Flows. These use a series of bijective (and therefore invertible) functions to transform from simple known densities (e.g. the unit normal) to the target density (\citealt{normalizing_flows};~\citealt{kingma2016improved}). MAFs represent $q$ as a transformation of a unit normal through a series of autoregressive functions (\citealt{MAF};~\citealt{MAF2}).

Masked Autoencoders for Distribution Estimation (MADEs,~\citealt{MADE}) are autoregressive density estimators as they parametrize the estimate in terms of one-dimensional conditionals. The density is factorized as

\begin{equation}
    p( \mathbf{d} | \boldsymbol{\theta}) = \prod_i^{\mathrm{dim} \big( \mathbf{d} \big) }  p( d_i | d_{1:i-1} , \boldsymbol{\theta}) \ \ 
\end{equation}

\noindent by masking the weights in the neural network. In this way the factorized probability for $d_1$ may depend only on $\boldsymbol{\theta}$, $d_2$ may depend on $d_1$ and $\boldsymbol{\theta}$, $d_3$ may depend on $d_1$, $d_2$ and $\boldsymbol{\theta}$, and so on. In a MADE, each estimated conditional is modelled as a Gaussian whose mean and variance are functions of $( d_{1:i-1}, \ \boldsymbol{\theta} )$ and are given by the neural network. The resulting function is a transformation to the target distribution from a space of random samples distributed according to a unit normal, where the associated Jacobian is triangular (due to the autoregressive structure) so can be easily calculated.

MAFs are composed of a series of MADEs, where the output of the last MADE is the input for the next. This allows for the estimation of more complicated densities that are not able to be factorized into a simple product of Gaussians (which the MADE requires). The repeated MADE layers in the MAF also allow the order of factorization to be shuffled, to better estimate general densities. 

As shown in Section~\ref{sec:results}, we use an ensemble of different network architectures for both MDNs and MAFs to validate the density estimation. The final density estimation is a stack of the ensemble estimates, weighted by the loss evaluated during training (see~\citealt{delfi2} for more details).

\subsection{Summary statistic compression} \label{sec:compression_intro}

For a given number of simulated mock data sets, density estimation likelihood-free inference is exponentially more efficient the lower the dimensionality of the data vector (i.e. summary statistics) $\mathbf{d}$. For likelihood-free inference to be scalable when forward simulations are expensive, it is typically necessary to compress high-dimensional data vectors down to some informative low-dimensional summaries $\mathbf{t}$. 

To this end, we want some compression function $\mathbf{t} = F(\mathbf{d})$, that is as informative as possible with respect to the unknown parameters whilst being as low-dimensional as possible. Under certain conditions, we can find a compression of a given data vector $\mathbf{d}$ down to $\mathbf{t}$ with dimension matching the number unknown parameters, $ {\mathrm{dim}}(\mathbf{t}) ={\mathrm{dim}}(\boldsymbol{\theta})$, that is lossless at the level of the Fisher information \citep{alsing_compression}.

Different approaches exist to try and achieve this compression, which are discussed in Section~\ref{sec:compression}. Crucially though, a poor choice of compression scheme (which we of course try to avoid) would lead to less informative summaries, but not to biased results. Provided the same compression scheme is applied to both the observed data and simulated mocks in a self-consistent way, subsequent likelihood-free inference of the parameters will be unbiased. A poor compression will however lead to added scatter in the forward-modelled samples $\mathbf{t}$ for a given set of parameters $\boldsymbol{\theta}$, leading to inflated parameter constraints.

In this work, our results use neural compression; using a neural network to learn the compression function. In particular, for the power spectrum and peak count summary statistics we use a regression network and for the deep CNN map compression we use both regression- and information- based training strategies for the network (see Sections~\ref{sec:compression} and ~\ref{sec:cnn_result} for details). We note that alternatively, neural compression can be achieved by \textit{information maximising neural networks} (IMNN), which aim to maximise the Fisher information~\citep{imnn}. We do not adopt the IMNN framework for this work however, as it requires specific simulations allowing for finite differences estimates of the gradients of summaries with respect to unknown parameters.

In this work we take two approaches to neural compression. In the first approach, we initially compress the mass maps down to some ``first level summaries" (in this case, power spectra and peak counts), which we then feed into a simple dense neural network for subsequent (massive) compression. In the second approach, we compress the map directly down to some informative low-dimensional summaries using deep convolutional neural networks. In Section~\ref{sec:cnn_intro} we introduce the details of our deep CNN map compression and show the likelihood-free results with the DES SV data in Section~\ref{sec:cnn_result}.

\section{Weak gravitational lensing} \label{sec:lensing}
Weak gravitational lensing is one of the foremost probes of cosmological large-scale structure. By using measurements of the galaxy shapes distorted by foreground matter due to gravitational lensing, we can directly infer density fluctuations in the total foreground matter (including non-visible dark matter). For convenience, here we have summarized some of relevant literature for weak gravitational lensing (see \citealt{bartelmann_schneider} and \citealt{kilbinger_cosmic_shear}). 

The weak lensing convergence $\kappa$ is given by a weighted projection of the density along the line of sight from the observer to a point with radial comoving distance $\chi$ and angular position $\vv{\phi}$ on the sky

\begin{equation} \label{eq:Q}
\kappa(\vv{\phi}, \chi ) = \frac{3 H_0^2 \Omega_m}{2} \int_0^\chi  \frac{\chi' (\chi - \chi')}{\chi} \frac{\delta(\vv{\phi}, \chi')}{a(\chi')} \ \mathrm{d} \chi' \  .
\end{equation}

\noindent where $H_0$ is the present value of the Hubble parameter, $a$ is the cosmological scale factor, $\Omega_m$ is the matter density parameter, $\delta$ is the overdensity, and the speed of light $c=1$. We have assumed flatness, such that the cosmological global curvature is zero, $K=0$.

For a radial (redshift) distribution $n(\chi)$ of lensed source galaxies, the convergence is given by

\begin{equation} \label{eq:Q2}
\begin{split}
\kappa (\vv{\phi}) &= \int_0^\infty n(\chi) \kappa(\vv{\phi}, \chi ) \mathrm{d} \chi \\
 &= \frac{3 H_0^2 \Omega_m}{2} \int_0^\infty \mathrm{d} \chi' f(\chi')  \chi' \frac{\delta(\vv{\phi}, \chi')}{a(\chi')} \ \ ,
\end{split}
\end{equation}

\noindent where 

\begin{equation} \label{eq:lensing_efficiency}
f(\chi') = \int^{\infty}_{\chi^\prime}  \left( \frac{\chi - \chi^\prime}{\chi}\right)n(\chi) \mathrm{d} \chi  \ .
\end{equation}

\noindent The convergence for the distribution of source galaxies at angular position $\vv{\phi}$ on the sky is therefore given by

\begin{equation} \label{eq:kappa_projected}
\kappa(\vv{\phi}) = \frac{3 H_0^2 \Omega_m}{2} \int_0^\infty  \Big[ \int_0^\chi\frac{\chi' (\chi - \chi')}{\chi} \frac{\delta(\vv{\phi}, \chi')}{a(\chi')}  \mathrm{d} \chi'  \Big] n(\chi) \mathrm{d} \chi \   \ .
\end{equation}

The shear field $\gamma$ is a spin-2 (\citealt{newman_penrose};~\citealt{wallis_projection}) field on the celestial sphere and is related to the convergence field through the lensing potential\footnote{See \citealt{bartelmann_schneider} and \citealt{kilbinger_cosmic_shear} for full details.} $\psi$:

\begin{equation}
\label{eq:kappa}
  \kappa = \frac{1}{4} (\eth \bar{\eth} +\bar{\eth} \eth) \psi \ \  
\end{equation}

\begin{equation}
\label{eq:gamma}
  \gamma = \frac{1}{2} \eth \eth  \psi \ \ ,
\end{equation}

\noindent where the differential operators $\eth$ and $\bar{\eth}$ are spin-weight linear operators~\citep{castro_heavens} defined on the sphere. It is therefore possible to determine the full-sky shear field $\gamma$ from the scalar convergence $\kappa$ (up to a constant of integration), for example by use of spin-weight spherical harmonic transforms. In this work, such transformations were used during the creation of the ideal shear $\gamma$ fields from convergence $\kappa$ fields (defined on the sphere) derived from simulations (see Section~\ref{sec:simulations}).

In the weak lensing limit, the observed ellipticity of a galaxy $\epsilon_{obs}$ is composed of both the intrinsic ellipiticity of the source galaxy $\epsilon_s$ plus the gravitational lensing shear $\gamma$. We therefore treat the measured ellipticity of a galaxy as an estimator for the  ``shear'', where the measurement is degraded by ``shape noise'' caused by the intrinsic ellipticity. In matrix notation, we can express a linear model with a data vector of observed shear measurements 

\begin{equation} \label{eq:matrix}
\boldsymbol{\gamma} = \boldsymbol{A} \boldsymbol{\kappa} + \mathbf{n} \ ,
\end{equation}

\noindent where $\mathbf{A}$ is the linear transformation between shear $\boldsymbol{\gamma}$ and convergence $\boldsymbol{\kappa}$ and $\mathbf{n}$ is a vector of noise per pixel. In our formulation the elements of the data vector are the galaxy shear measurements binned into pixels depending on their sky position.

\begin{figure}
\hspace*{-0.5cm}
\includegraphics[width=0.51\textwidth]{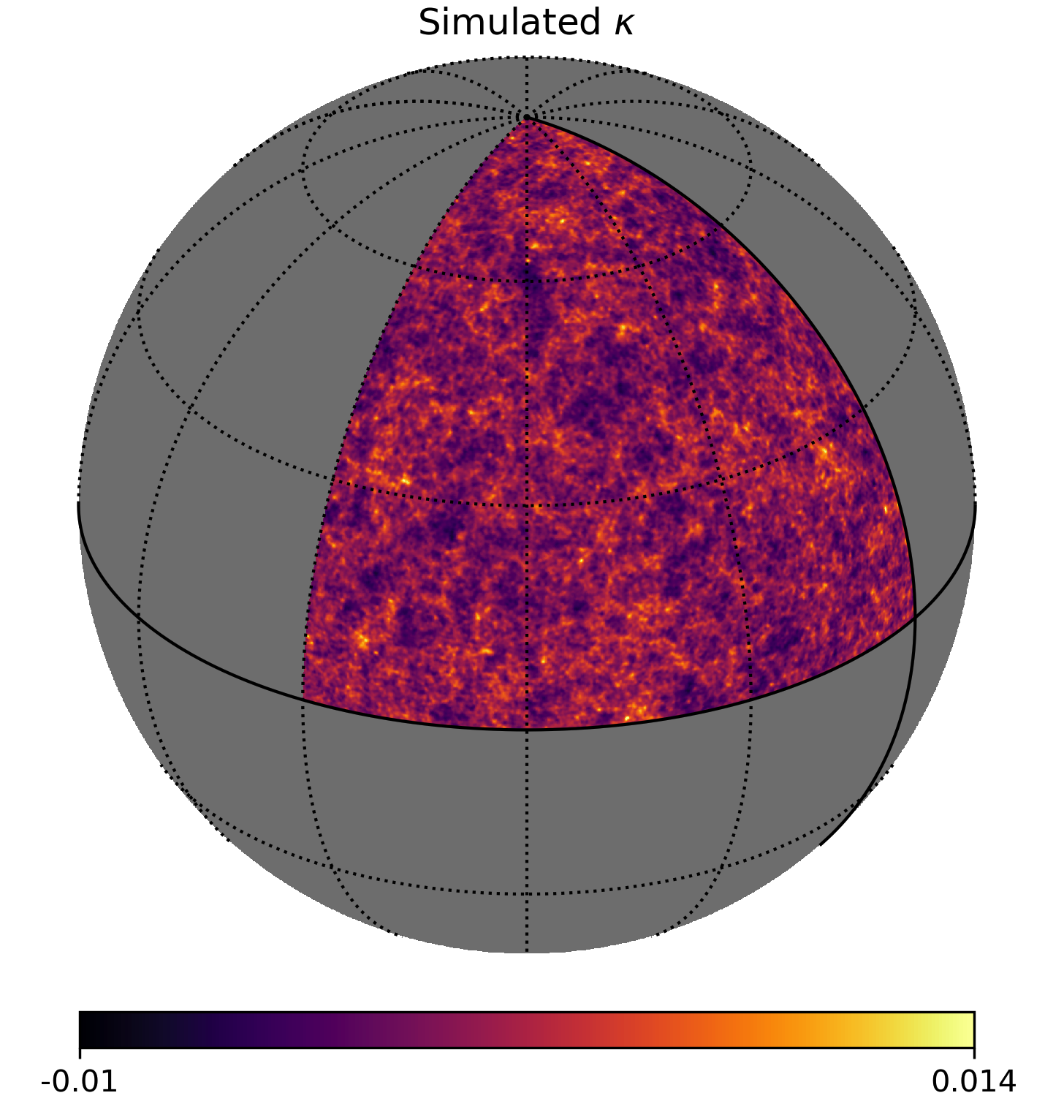}
\caption{\label{fig:kappa_picola} Example octant convergence $\kappa$ field calculated by ray tracing a {\sc L-Picola} dark matter simulation. From such a field we generate 18 non-overlapping DES SV mock data sets. The convergence field has been smoothed with a $\sigma =10$ arcmin Gaussian kernel for visualization purposes.}
\end{figure}

\section{Summary statistics} \label{sec:statistics}
\subsection{Kaiser-Squires map reconstruction} \label{sec:ks}
For smaller patches of sky, in the \textit{flat-sky approximation} the $\eth$ operators on the sphere may be reduced to $\partial$ derivatives with respect to the two sky angles $\phi_1$ and $\phi_2$, so that shear $\gamma$ may be related to convergence $\kappa$ as

\begin{equation} \label{eq:flat_sky}
\tilde{\gamma} (\mathbf{k}) = \frac{k_1^2 -k_2^2 + 2 i k_1 k_2}{k_1^2 + k_2^2} \tilde{\kappa} (\mathbf{k})  \  \  ,
\end{equation} 

\noindent where $k_1$ and $k_2$ are elements of the two-dimensional Fourier vector $\mathbf{k}$, so that 

\begin{equation}
\begin{split}
\gamma( \vv{\phi} ) = \frac{1}{\pi} \int_{\mathbb{R}^2} \mathrm{d}^2 \phi' {\Gamma} ( \vv{\phi} - \vv{\phi}') \kappa(\vv{\phi}') \\
\mathrm{where} \; \Gamma  ( \vv{\phi} ) = - ( \phi_1 - i \phi_2 ) ^{-2} \ .
\end{split}
\end{equation}

\noindent The~\cite{ks93} reconstruction method uses the pixelized observed $\boldsymbol{\gamma}$ map to reconstruct the unknown $\boldsymbol{\kappa}$ by inverting equation~\ref{eq:flat_sky}. This procedure to infer $\boldsymbol{\kappa}$ takes no account of varying shape noise in the shear map or masks, which introduce artefacts into the recovered convergence \textit{mass map}. Figure~\ref{fig:ks} shows the reconstruction for the DES SV data (as described in Section~\ref{sec:simulations}).

The Kaiser-Squires reconstruction, in addition to not accounting for spatially varying noise, includes no explicit prior information about the signal.\footnote{Implicitly, the reconstruction is the {\textit{maximum a posteriori}} estimate under the assumption of Gaussian noise and a uniform prior $p(\kappa)$.} This will not lead to incorrect inferences in the likelihood-free inference framework, as the anisotropic noise and mask will be forward modelled in the mock simulated data. Different reconstruction methods that use prior information about the signal have been shown to more accurately reconstruct the convergence field $\kappa$, either in closed-form (\citealt{marshall_mass_maps};~\citealt{glimpse2016};~\citealt{alsing2016hierarchical};~\citealt{alsing2017cosmological};~\citealt{jeffrey2018};~\citealt{price_maps}) or implicitly learned using samples from the prior (\citealt{deeplearning_shirasaki};~\citealt{deepmass}). More accurate mapping methods would likely increase the signal-to-noise of summary statistics, and therefore improve constraining power, and a study of this in the context of likelihood-free inference would merit future work.

\begin{figure}
\hspace*{-0.5cm}
\includegraphics[width=0.51\textwidth]{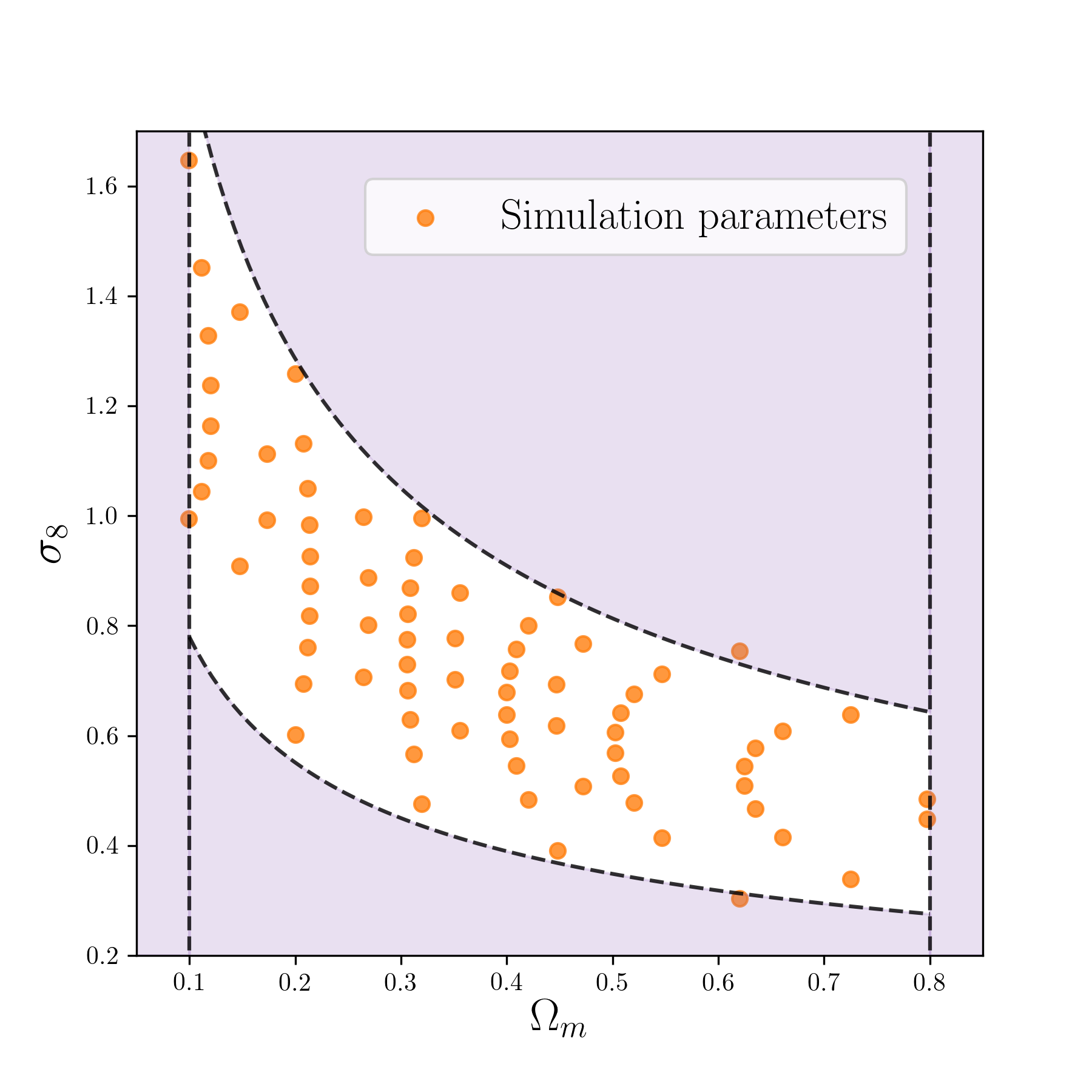}
\caption{\label{fig:sim_points} The 74 parameter pairs corresponding to the input cosmology of the forward-modelled data. The dashed lines shows the limits of the prior range (as described in Section~\ref{sec:results}) and the shaded region is therefore excluded by the prior during parameter inference.}
\end{figure} 

\begin{figure*}
\centering
\includegraphics[width=0.7\textwidth]{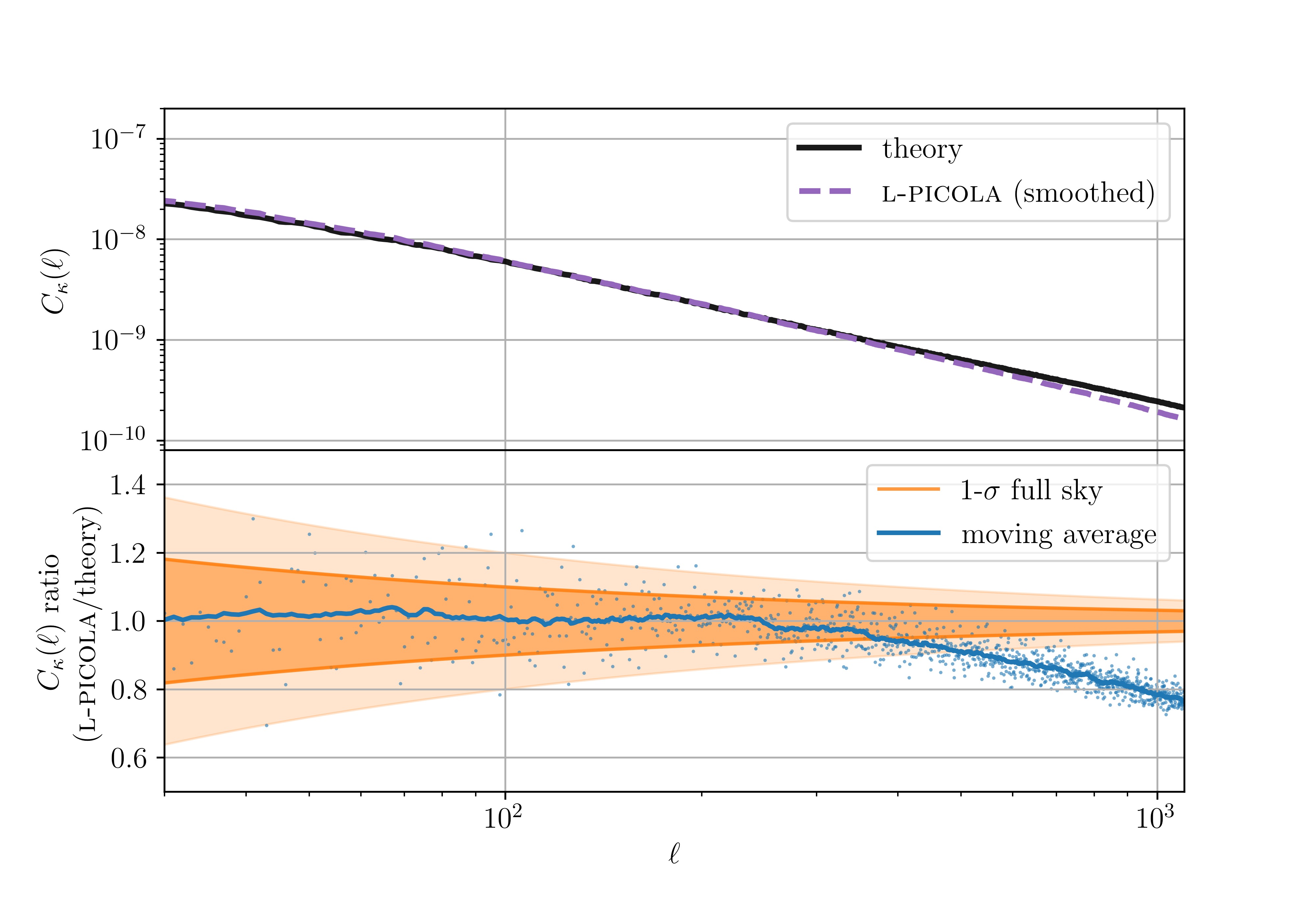}
\caption{\label{fig:lensing_power} Example of a full-sky power spectrum of convergence $C_\kappa(\ell)$ from theory ({\textsc{Nicaea}}) and a single L-PICOLA simulation realization. The shaded 1- and 2-$\sigma$ regions correspond to full-sky measurement uncertainty due to cosmic variance. For our case, the sky fraction of the observed data is less than $5 \times 10^{-3}$; the resulting sample variance will in fact give a confidence region that is a factor of more than 10 greater than the confidence region from full-sky cosmic variance alone. Nevertheless, we still correct the power spectra of the simulated convergence fields by rescaling the spherical harmonic coefficients $a_{\ell m}$ as described in Section~\ref{sec:simulations}.}
\end{figure*} 

\subsection{Power spectrum}

The angular power spectrum for the convergence field $\kappa$ on the celestial sphere is given by

\begin{equation}
  \langle a_{\ell  m} a_{\ell'  m'}^* \rangle =  C_\kappa (\ell) \delta_{m m'} \delta_{\ell \ell'} \ \ ,
\end{equation}

\noindent with spherical harmonic coefficients $a_{\ell  m}$ of the convergence $\kappa$ field, where we have used Kronecker delta $\delta_{m m'}$. The expectation, $\langle \rangle$, is with respect to random realizations. 

For a given field on the sphere, an unbiased estimate of the power spectrum is given by 

\begin{equation}
    \hat{C}_\kappa (\ell) = \frac{1}{2 \ell +1} \sum_{m=-\ell}^{m=+\ell} | a_{\ell m}  a_{\ell m}^*| \ \ .
\end{equation}

For a simple contiguous masked region (e.g. an octant of the sky) the measured power spectrum can be rescaled by the fraction of sky observed $f_{\rm sky}$ to give an approximate unbiased estimate; in this approximation $ C_{\ell}  \approx \langle f_{\rm sky} \hat{C}_{\ell} \rangle$~\citep{Dodelson}. The variance due to cosmic variance (i.e. finite $m$ modes per $\ell$) and additional sample variance due to finite sky coverage is then given by

\begin{equation} \label{eq:cosmic_variance}
    \sigma_{C_\ell}^2 = \frac{1}{f_{\rm sky}} \frac{2}{2 \ell + 1}  \  C_\ell^2 \ .
\end{equation}

Figure~\ref{fig:lensing_power} shows the theoretical convergence power spectrum and the power spectrum measured from the {\textsc{L-Picola}} simulations~\citep{picola}. The source of the discrepancy between the two power spectra is discussed further in Section~\ref{sec:mock_shear}. However, the 1- and 2-$\sigma$ confidence intervals in Figure~\ref{fig:lensing_power} correspond to a full sky cosmic variance uncertainty. For the 139 sq deg area of the DES SV data, the actual 1-$\sigma$ region would be a factor of more than 10 greater.

As the data cover a small sky area we do not use spherical harmonics (similarly to the map-making). Instead we measure the power spectrum of our complex KS maps (i.e. combining the E- and B-modes) using Fast Fourier Transforms (FFT, ~\citealt{fft}). We bin the power spectrum into 18 Fourier band powers, with centres of  $ k_{\rm centre}   / \big( 10^{-3}   {\rm arcmin}^{-1} \big ) $:

\begin{center}
 \begin{tabular}{||c c c c c c  ||} 
 \hline
 0.434 &  1.183 &  2.044 &  2.997 &  3.939 &  5.177 \\ 
 \hline 
  6.805  & 8.943 &  11.75 &  15.44 &  20.30 &  26.68  \\
 \hline 
 35.07 & 46.10 &  60.59 &  79.63 &  104.6 &  137.5 \\ [1ex] 
 \hline
\end{tabular}
\end{center}

\noindent Since we are in the likelihood-free framework, we do not need to correct for this flat-sky approximation, as we perform the same operations self-consistently to the simulated mock data and the actual observed data.

Equivalently with the mask and noise in the power spectrum, we measure the ``raw'' power spectrum from the maps in the same manner for both the simulated mock data and the actual observed data. Subtracting and rescaling the measured spectra to account for the mask and noise effects, essentially the same as a scalar \textit{pseudo}-$C_\ell$ estimate (PCL,~\citealt{pcl}), is implicitly taken care of during the likelihood-free inference.

In this work we have measured the unsmoothed Kaiser-Squires map $\kappa_{\rm KS} = \kappa_{\rm KS,\ E} + i \kappa_{\rm KS,\ B} $. This was found to perform marginally better with simulations than the $\kappa_{\rm KS,\ E}$ map alone. One could measure the power spectrum of both separately and combine them, thus keeping the maximum amount of information; this would increase the size of the summary statistic data vector, making the compression step more difficult, yet with little added constraining power. It is therefore possible that a spin \textit{pseudo}-$C_\ell$ estimate of the E-mode power using both contributions would provide a more constraining summary statistic~\citep{taylor_delfi_lensing}, while keeping the size of the to-be-compressed data vector low. Similarly, a minimum-variance power spectrum estimate (e.g. \citealt{tegmark97}) would provide the optimally-constraining power spectrum without increasing the size of the data vector in likelihood-free inference.

\subsection{Peak statistics} \label{sec:peaks}

For an isotropic Gaussian random field, the mean (zero by definition for our $\kappa$ field) and the power spectrum are sufficient to entirely characterise the field. Of course, in the late-Universe, non-Gaussianity arises due to non-linear structure formation, for which summary statistics beyond the power spectrum are suited to provide additional information to constrain cosmological parameters.

Counts of the number of peaks in a mass map are particularly promising, as peaks in the density field probe the non-Gaussian structure directly in a manner that is sensitive to the changes in the cosmological parameters (\citealt{dietrich_peaks};~\citealt{kids_peaks1};~\citealt{kids_peaks2};~\citealt{peel_peaks}). In previous work,~\cite{peaks_lin} used a likelihood-free inference method, the ABC rejection sampling technique (rather than density estimation as in this work) applied to simulated data, in which the forward model was a fast halo approximation and, therefore, can not easily be used for joint power spectrum and peak constraints without considerable adaptation.

We define our peak summary statistic as the number of pixels in the smoothed Kaiser-Squires map reconstruction that are of a value greater than all of their neighbours, which we bin according the convergence $\kappa$ value of the peak, $n(\kappa)$. The smoothing scale of $10\ {\rm arcmin}$ was shown in~\cite{jeffrey2018} to give a map reconstruction closest to the underlying convergence $\kappa$ field.

As discussed in Section~\ref{sec:ks}, there are reconstruction methods beyond Kaiser-Squires that can be more accurate and would therefore likely give more informative peak statistic summaries, and these deserve future investigation. The choice of a given reconstruction method with a certain smoothing scale and definition of peak corresponds to a specific choice of summary statistic. The peak statistic analysis in this work is, therefore, not equivalent to~\cite{des_sv_peaks}, where different choices and definitions were used.

\begin{figure*}
\centering
\includegraphics[width=0.99\textwidth]{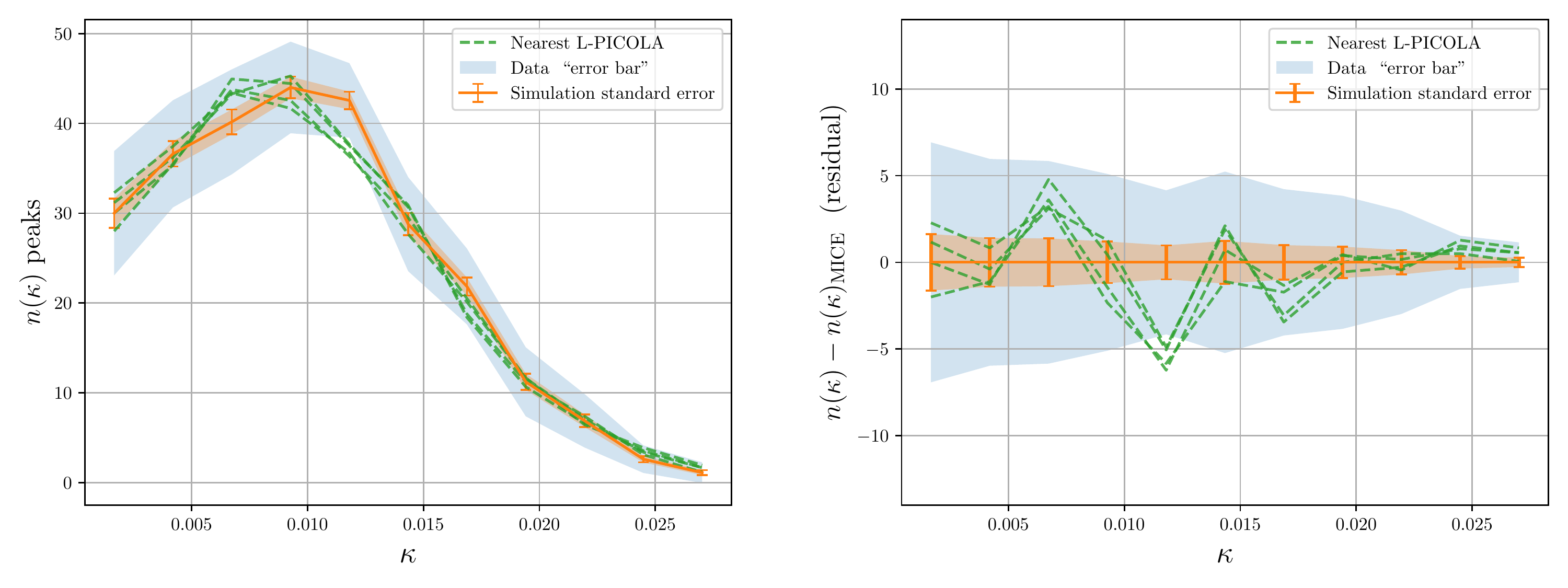}
\caption{\label{fig:mice_validation} Validation of peak count summary statistic prediction from {\textsc{L-Picola}} simulations with respect to the prediction from MICE simulations. This shows the mean predicted {\textsc{L-Picola}} peak statistic from four points in parameter space closest to the MICE cosmological parameters. The {\textit{left panel}} shows the absolute peak count summary statistic and the {\textit{right panel}} shows the difference between {\textsc{L-Picola}} and MICE. Given the noise level of this data (corresponding to the larger blue confidence interval), the discrepancies will not lead to a significant parameter shift, though the discrepancy could have some impact for a key result from a cosmological survey.}
\end{figure*}

Figure~\ref{fig:mice_validation} shows the peak statistics measured from simulated mock data, showing our 11 bins between $\kappa=0$ and $\kappa=0.028$. It was noted by~\cite{kids_peaks2} that adding peaks with negative $\kappa$ did not lead to more informative summaries, ostensibly because they are strongly correlated with high-valued peaks.  The validation of our simulated peak statistic summaries, as shown in Figure~\ref{fig:mice_validation}, is discussed in Section~\ref{sec:simulations}.

\subsection{Deep convolutional features} \label{sec:cnn_intro}

As another way to access the non-Gaussian information (beyond two-point) of the weak lensing field, deep convolutional networks have recently attracted significant attention. Instead of relying on crafted non-Gaussian statistics, like peak counts, which are based on a priori understanding of the physics, CNNs can be seen as flexible non-linear feature extractors; they can be optimized as to find maximally relevant summary statistics from the mass map. CNN outputs have been first applied as a cosmological summary statistic by \cite{fluri_cnn},~\cite{ribli_cnn}, and~\cite{kids_cnn}.

CNNs are particularly suited for two-dimensional image or one-dimensional time series data with translation invariant features in the underlying signal. Mathematically, they are a sequence of iteratively computed layers. At a given layer $j$ the signal $\mathbf{x}_j$ is computed from the previous layer 

\begin{equation}
\mathbf{x}_j = \rho \mathbf{M}_j \mathbf{x}_{j-1}
\end{equation}

\noindent with linear operators (i.e. convolutions) $\mathbf{M}_j$ and nonlinear {\it activation function} $\rho$  (\citealt{lecun1990handwritten};~\citealt{mallat2016understanding}). Deep architectures, with a series of additional layers, are often able to learn features with greater complexity than shallow architectures. For a general overview of deep learning and neural networks see~\cite{deep_learning}. 

Recent works have demonstrated that the statistics extracted by a CNN are more powerful than both the two-point functions and peak counts, hinting that such models are accessing additional cosmological information present in the data. 

This should not be surprising as CNNs have the capacity to be sensitive to both global scales and local features, and as such, should be able to be sensitive to both the power spectrum and peak count signals (\citealt{ribli_cnn};~\citealt{wst_cnn}).

In contrast to most other works which adopt simple CNN architectures, in this work we opt for a state-of-the-art deep residual network model \citep{resnet}. In recent years \textit{ResNets} have become an established standard architecture for image classification and regression tasks, significantly outperforming simpler CNNs~\citep{he2016identity}. 

Our ResNet model, which we will refer to as a \textit{deep compressor}, accepts the noisy Kaiser-Squires map as an input, and is tasked with returning two numbers, which will constitute our summary statistics. The details of the model are presented in Section~\ref{sec:cnn_result}.

\section{Data and forward model} \label{sec:simulations}

\subsection{DES SV data}

DES is a ground-based photometric galaxy survey, which observed in the southern sky from the 4m Blanco telescope in Chile with five photometric filters~\citep{decam}. The SV (A1) data\footnote{\url{http://des.ncsa.illinois.edu}} come from an initial run of 139 deg$^2$, but with depth approximately that of the full 6 year survey~\citep{chang_sv_map}. Data selection choices match \cite{jeffrey2018}.

We make a redshift cut of $0.6<z_{\rm mean}<1.2$, where $z_{\rm mean}$ is the mean of the $z$ posterior for each galaxy. In our analysis we use a single tomographic redshift bin for all selections, matching the shear peak analysis for this data set performed in~\citealt{des_sv_peaks}. By using more bins, one could generate multiple maps that probed different redshifts through a range of peak lensing kernels, giving the possibility of more constraining power.

\subsection{Dark matter simulations}

We use 74 independent dark matter simulations, each with a different pair of cosmological parameters ${\Omega_m}$ and $\sigma_8$, and each covering an octant of the celestial sphere. All simulations used a standard flat $\Lambda$CDM cosmological model with Hubble parameter $H_0=70\ {\mathrm{km \ Mpc^{-1} s^{-1}}}$ and fixed values of the scalar spectral index and baryon density,  $n_s=0.95$ and $\Omega_b =0.044$ respectively. Fixing these parameters can be interpreted as using a Dirac delta as the prior probability distribution for these parameters during inference.

The dark matter simulations are generated using the {\sc l-picola} code~\citep{picola}, which is based on the {\sc cola}~\citep{cola} algorithm. This uses a combination of second-order Lagrangian perturbation theory (2LPT) and a Particle-Mesh (PM) which requires fewer time steps than ``full'' $N$-body (e.g. Gadget~\citealt{gadget}) and which therefore can generate simulations more quickly.  

We used a 1250 Mpc$/h$ comoving simulation box, $768^3$ particles, and a $1536^3$ grid. A $z<1.6$ lightcone was generated with up to four box replicates, using 30 time steps from $z=20$. The initial conditions used the linear matter power spectrum from~\citealt{eisenstein_hu}.
 
\subsubsection{Mock shear maps} \label{sec:mock_shear}

To generate a convergence map from a simulation, the matter particles were binned using the {\sc healp}ix~\citep{healpix} pixelization of the sphere with {\sc nside}=2048 in comoving radial shells of $50\ {\rm Mpc/h}$. The particle density $\rho$ map in a given redshift was converted into an overdensity $\delta = \rho / \Bar{\rho} -1$ using the average density in the shell $\Bar{\rho}$. The convergence was calculated per pixel using equation~\ref{eq:kappa_projected}, under the Born approximation (see Appendix~\ref{append:codes}). We use the $n(z)$ in the lensing kernel matching the DES SV data, which we approximate by summing the individual posterior redshift distributions per galaxy from the BPZ photometric redshift code \citep{bpz2}, matching the original analysis of this dataset~\citep{shear_des_sv}. The convergence maps were downgraded to {\sc nside}=1024. Figure~\ref{fig:kappa_picola} shows an example of the resulting convergence map.

The drawback of the COLA approach is the accuracy of the dark matter distribution. The finite spatial resolution and fewer timesteps used by the {\sc cola} method particularly affect small distance scales. Figure~\ref{fig:lensing_power} shows a suppression of the {\sc l-picola} power spectrum at scales of $\ell > 700$ of order 10 per cent (relative to {\sc Nicaea}\footnote{\url{nicaea.readthedocs.io}}~\citep{nicaea} theory), as is expected with {\sc cola} methods. We correct the power of the {\sc l-picola} convergence using the~\textsc{Nicaea} predictions with halofit~\citep{halofit}. 

We estimate the smooth part of the $C_\kappa (\ell)$ from a convergence map realization using a polynomial order 1 Savgol filter with window size 91 for each convergence map and reweighting spherical harmonics. We rescale the spherical harmonic coefficients of the map by the ratio of the {\sc nicaea} theory and only the smooth part of the measured simulation power spectrum, which ensures that the natural fluctuations inherent in $C_\kappa (\ell)$ for a given realization are preserved.

The octant convergence fields are transformed to shear fields using spherical transformations as described in Section~\ref{sec:lensing}. The mask introduces small errors in the large scale shear field, which are negligible for the much smaller data patches, especially once the non-smooth data mask and noise are included. 

To generate mock DES SV shear maps, square patches of 256$\times$256 pixels with 4.5 arcmin resolution are extracted from the spherical map with a gnonomic projection. The 139 square deg DES SV mask is then applied by excluding pixel data outside the observed area. From an octant we extract 18 non-overlapping DES SV mock data sets.

\begin{figure*}
\centering 
\includegraphics[width=1.02\textwidth]{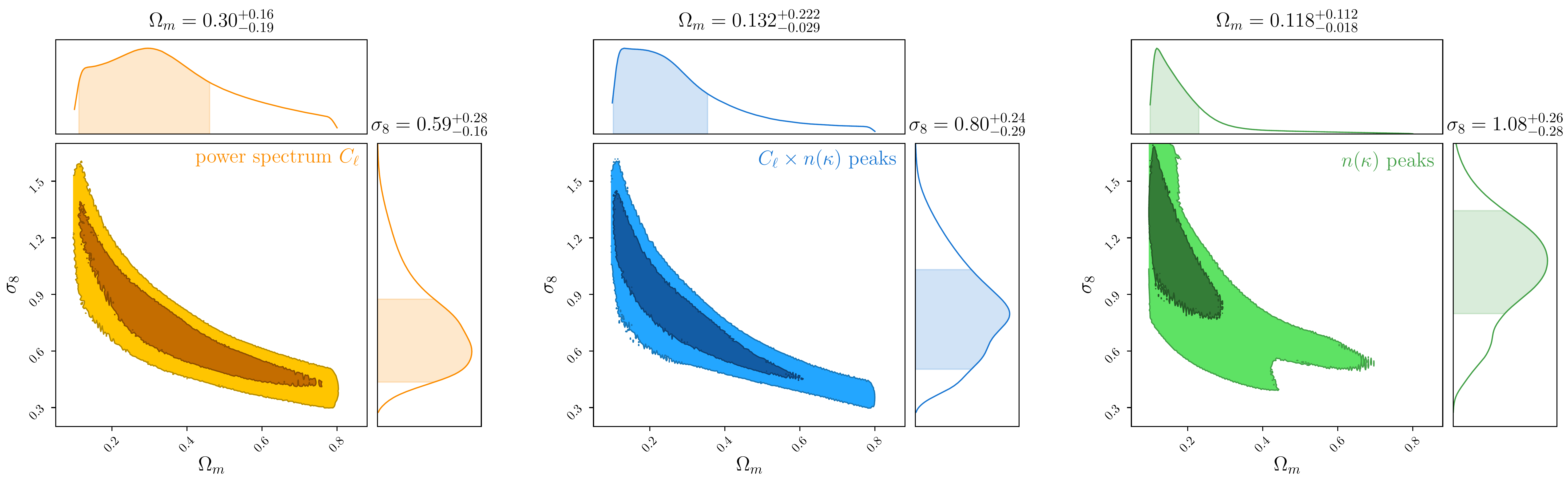}
\caption{\label{fig:joint_post} Posterior probability distributions from neural compressed summary statistics. The {\textit{left panel}} shows the posterior probability distribution from the power spectrum, the  {\textit{centre panel}} shows the joint peaks and power spectrum, and the {\textit{right panel}} shows the posterior from peaks alone. The resulting marginal posterior distribution for $\Omega_m$ from peaks alone is particularly peaked due to the prior lower bound for $\Omega_m$, not due to the data constraints.}
\end{figure*} 

Noise realizations are generated from the data by randomly exchanging the ellipticity values between galaxies in the catalogue to remove the lensing signal and leave the shape noise remaining; this standard process assumes the variance due to lensing is negligible contribution to the ellipticity variance per galaxy. This method generates realistic, non-Gaussian shape noise in our mock catalogues.

Of the effects excluded from our forward modelling, the uncertainty in the redshift distribution and the intrinsic alignments of galaxies~\citep{kirk_ia} are likely to have the largest impact. In future work, these effects can be included in the mock data with a chosen data model, and any associated nuisance parameters included in the inference (for treatment of nuisance parameters while keeping low-dimensional summary statistics for likelihood-free analyses, see \citealp{alsing2019nuisance}).

\subsubsection{Peak count validation}

By construction the power spectra of the mock data match those predicted by theory (see Section~\ref{sec:mock_shear}). For the peak statistics, we validate our mock data against higher-resolution and more accurate $N$-body predictions.

We use the public MICE lensing simulations~\citep{mice3} as a higher resolution ``truth'', against which we can validate our forward-modelled data that used the approximate {\textsc{L-Picola}} $N$-body method. We generate 18 non-overlapping MICE maps, using the same source galaxy distribution, mask and shape noise generation method, to match the {\textsc{L-Picola}} mock data. From these mock maps the peak count summary statistics are calculated identically in both cases.

All cosmological parameters in our forward-model pipeline, other than those to be inferred from the data, are fixed to the MICE values. We take four points in parameter space $[ \Omega_m, \sigma_8 ]$ closest to the MICE values, such that the four points form the corners of a quadrilateral with the MICE parameter values inside. At each of these four parameter points, we take the average of the 18 mock peak count summary statistics; in Figure~\ref{fig:mice_validation} these are labelled `Nearest L-PICOLA'.

From the 18 MICE realizations, we calculate the mean and the standard deviation for each element. The standard deviation corresponds to the larger blue shaded confidence interval in Figure~\ref{fig:mice_validation} labelled `Data ``errorbar'''. This standard deviation corresponds to the square root of the diagonal of the data covariance (as used in a standard likelihood analysis), and so tells us whether a discrepancy between MICE and {\textsc{L-Picola}} is larger than the expected scatter in the observed data. The smaller orange confidence interval with plotted errorbars are the standard errors on the mean (the larger confidence interval scaled by $18^{-\frac{1}{2}}$). In the left panel of Figure~\ref{fig:mice_validation}, there is agreement with the overall shape of the peak count distribution between MICE and {\textsc{L-Picola}}. The right panel shows the difference between {\textsc{L-Picola}} and MICE, in which shows that there are some differences between all four {\textsc{L-Picola}} curves for certain $\kappa$ bins, in particular the bins either side of the $n(\kappa)$ peak and the two highest $\kappa$ bins. 

To quantify the discrepancy and ascertain whether it is significant, we evaluate the reduced chi-squared statistic. As a conservative approximation, we take the mean of the four {\textsc{L-Picola}} curves as the prediction $\boldsymbol{\mu}$ and calculate the reduced chi-squared 

\begin{equation}
    \chi^2_\nu = \frac{1}{\nu} \sum_{i j} (d_i - \mu_i)^T \Sigma^{-1}_{i j} (d_j - \mu_j) \ \ ,
\end{equation}

\noindent for the different elements of the data vector $d_i$ with degrees-of-freedom $\nu = 11$ (as the unknown parameters are fixed) and covariance $\Sigma$ estimated from the four {\textsc{L-Picola}} sets (a total of $72$ realizations). 

For $\chi^2_\nu$ using the uncertainty due to the finite number of MICE simulations, we scale $\Sigma$ by $18$, resulting in $\chi_\nu \approx 3.4$. For the realistic data covariance, we calculate $\chi_\nu \approx 0.2$. There is therefore a measurable difference between MICE and {\textsc{L-Picola}} with a poor fit of $\chi_\nu > 1$ with simulation errors alone. However, with realistic data noise, a $\chi_\nu \ll 1$ implies that this discrepancy is subdominant in comparison to other uncertainty contributions.

The discrepancy would not give a significant parameter shift, as it is within the noise level of the data. Some shift is nevertheless still there, so this discrepancy may be too large for key results from cosmological surveys. Additionally, for data sets more constraining than the DES SV data, this discrepancy would become more significant.

By using a simulation with only dark matter particles to validate our forward-model, we are excluding tests for baryonic effects. These effects should appear at small scales, and we are likely to have suppressed them for DES data due to our $10\ {\rm arcmin}$ smoothing~\citep{baryonic_smoothing}. Further validation tests with high-resolution simulations with baryonic effects included would validate this.

\section{Results: power and peak summary statistics} \label{sec:results}

\subsection{Overview}

In this section we present the likelihood-free inference results using the power spectra and peak count summary statistics measured from the DES SV weak gravitational lensing mass maps. These are crafted summaries that we expect to be informative with respect to our unknown cosmological parameters $\boldsymbol{\theta}$. We leave the results from the deep compressor, extracted informative summaries from the lensing mass maps using deep CNNs, to the next section.

For the power and peak summary statistics, we first use a neural compressor to reduce our high-dimensional measured summaries $\mathbf{d}$ to low-dimensional compressed summaries $\mathbf{t}$. We then estimate the density $p(\mathbf{t}|\boldsymbol{\theta})$ using pyDELFI. We validate this density using an ensemble of neural density estimators.

The likelihood is then evaluated for the compressed summaries of the observed (compressed) data $\mathbf{t}_o$ and combined with the prior $p(\boldsymbol{\theta})$ to give our posterior probability for the parameters given our compressed power spectra and peak count summary statistics.

\begin{figure*}
\includegraphics[width=0.7\textwidth]{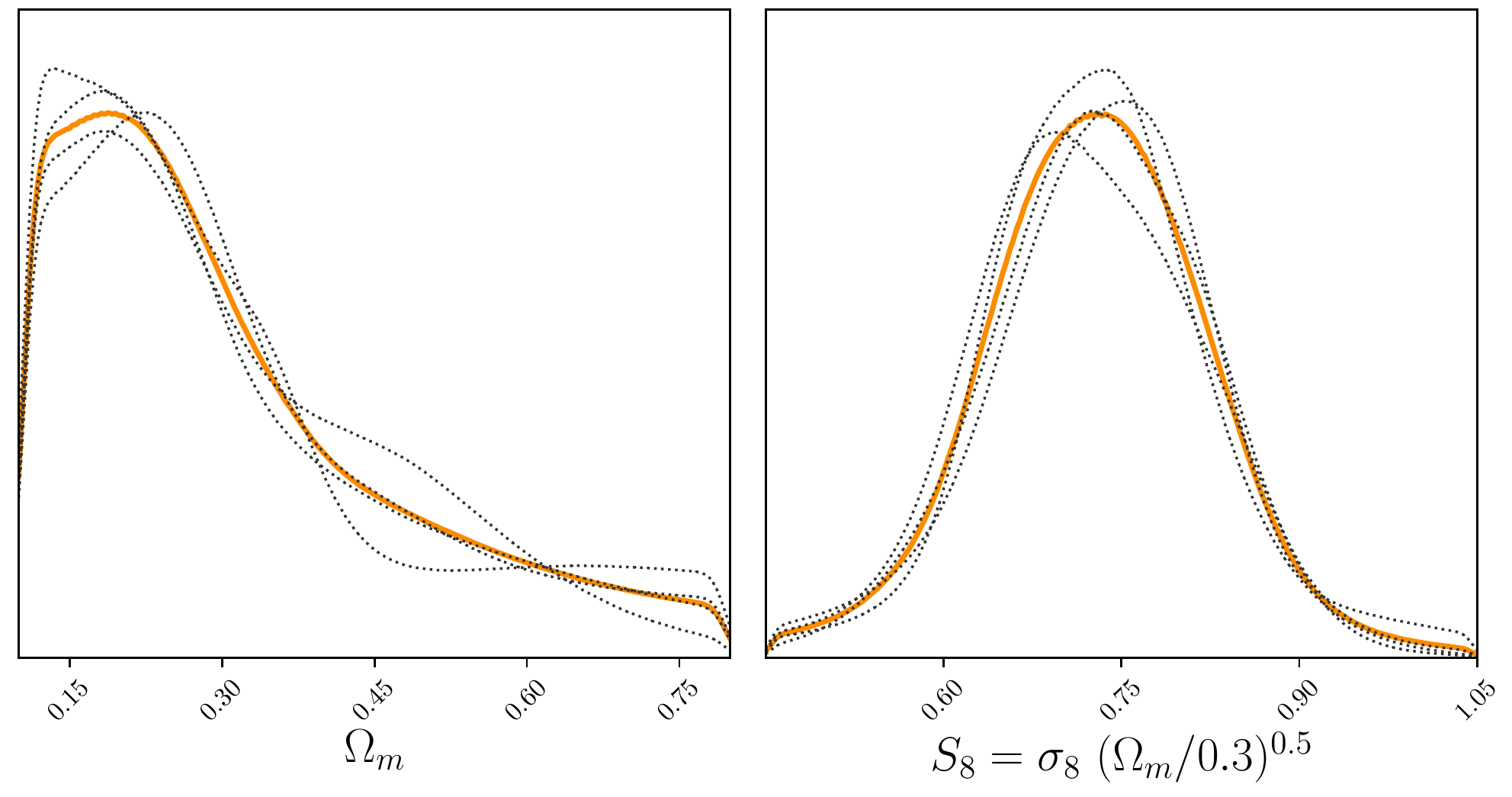}
\caption{\label{fig:marginals} Validation of probability density estimation: marginal posterior distributions for the different neural density estimators in the pyDELFI ensemble with the (neural compressed) joint peak and power spectrum data. The orange line is the marginal posterior using the final stacked density. Though the individual density estimates are stacked with weights corresponding to their loss during training, there is nevertheless general agreement between the estimates. In future surveys, these distributions would give an estimate for where new simulations should be run in parameter space during {\textit{active learning}}.}
\end{figure*} 

\subsection{Data compression} \label{sec:compression}

Using pyDELFI we aim to estimate the conditional distribution $p(\mathbf{t} | \boldsymbol{\theta})$, with compressed summary statistic $\mathbf{t}$. As discussed in Section~\ref{sec:compression_intro}, the compression of the summary statistics down to the same dimension as the parameters $\boldsymbol{\theta}$ (in this case two) is done to improve the density estimation with finite simulated summaries; the density estimation is done in $\{\mathbf{t}, \boldsymbol{\theta} \}$ space, rather than $\{\mathbf{d}, \boldsymbol{\theta} \}$.

In~\cite{alsing_compression} it was shown that for an underlying true likelihood, the optimal compression is lossless for a fiducial value of the parameter $\boldsymbol{\theta}$. For an underlying Gaussian likelihood, this \textit{score compression} is linear (assuming the covariance is parameter-independent) and corresponds to the {\textsc{Moped}} compression~\citep{moped} often used in astrophysical data compression. In general the optimal score compression reproduces summaries that are transformations of, and have the same constraining power as, the maximum likelihood estimate under the assumed likelihood parameter estimates, $\mathbf{t} = F(\mathbf{d}) = \boldsymbol{\theta}_{\mathrm{MLE}}$.

Rather than using an assumed likelihood, our main result uses neural compression, where $\mathbf{t} = F_\varphi (\mathbf{d})$ is approximated by a neural network. This is achieved by training over an augmented training set of $\{\mathbf{d}^+ , \boldsymbol{\theta}^+ \}$, aiming to minimise the following loss function with respect to the neural network parameters $\boldsymbol{\varphi}$:

\begin{equation} \label{eq:mse_loss}
J(\boldsymbol{\varphi}) = || F_{\boldsymbol{\varphi}} (\mathbf{d}^+) -  [ \mathbf{A} \boldsymbol{\theta}^+ + b ] ||^2_2 \ \ ,
\end{equation}

\noindent which is averaged over simulated data $\mathbf{d}^+$ and parameter pairs $\boldsymbol{\theta}^+ $ from the augmented training set, with fixed rescaling $\mathbf{A}$ and shift $b$ to normalize for efficient training depending on network architecture. In this approach, unlike for the density estimation, the mock data $\mathbf{d}^+$ need not be independent, and we therefore generate 2500 data realization per simulated convergence octant to create the augmented training set.

In comparison to the score compression having an equivalence with maximum likelihood estimation for an assumed likelihood, this choice of loss $J(\boldsymbol{\varphi})$ has an equivalence with a mean posterior estimate~\citep{jaynes07}, but without an assumed likelihood and with an implicit prior given by the distribution of the training labels $\boldsymbol{\theta}^+ $.

An alternative commonly used loss is the $L_1$ norm, e.g. $||F(Y) - X ||_1 $, corresponding to minimizing the least absolute deviation (LED) or mean absolute error (MAE), which would have an equivalence to a median posterior estimate. This was similarly tested, but resulted in more lossy compression when tested with simulated data (see Appendix~\ref{append:neural}).

However, once the neural compression is trained for a given summary statistic using the augmented set $\{\mathbf{d}^+ , \boldsymbol{\theta}^+ \}$, the compressor is fixed and the density estimation of $p(\mathbf{t} | \boldsymbol{\theta})$ is performed with pyDELFI with the unsimulated observed data. A sub-optimal data compression will lead to larger scatter in $\{\mathbf{t}, \boldsymbol{\theta} \}$ spaces, leading to inflated parameter constraints, but not to incorrect inference.\footnote{As with all data analysis, poor data compression or cuts can lead to loss of information and less ability to infer the unknown parameters.}

The chosen network architecture and training scheme are described in Appendix~\ref{append:neural}. 

\subsection[pyDELFI density estimation]{pyDELFI density estimation of $p(\mathbf{t} | \boldsymbol{\theta})$}

To robustly estimate the distributions $p(\mathbf{t} | \boldsymbol{\theta})$, we used an ensemble of neural density estimators with pyDELFI.

We used two (full-rank) Gaussian Mixture Density Networks (MDNs) and two Masked Autoregressive Flows (MAFs). The two MDNs had two and three Gaussian components respectively, and both had two dense hidden layers with 30 neurons per layer. The two MAFs had four and five MADE layers respectively, each with two dense hidden layers with 50 neurons per layer. We used tanh activation functions throughout. The final reconstructed density $p(\mathbf{t} | \boldsymbol{\theta})$ is then taken as a weighted average of the individual neural density estimators, weighted by the value of the loss achieved during training (i.e. model averaged relative to their individual performances).

The density estimation was carried out using the transformed parameters $\boldsymbol{\theta}' = [\Omega_m, S_8 =  \sigma_8 (\Omega_m /0.3)^{0.5}]$. This coordinate transformation simplifies the density estimation task, and the densities can then be transformed back to $\boldsymbol{\theta} = [\Omega_m, \sigma_8]$. We restrict the density estimation procedure to our eventual prior range of $0.1<\Omega_m < 0.8$ and $0.45 < S_8 < 1.05$.

Beyond the usual training-validation split during training implemented in pyDELFI, combined with early-stopping to avoid over-fitting, the individual estimates from the neural density ensemble can be used as a further (visual) validation step. Figure~\ref{fig:marginals} shows the marginal posterior probabilities (see next section for details) for the parameters $\Omega_m$ and $S_8 =  \sigma_8 (\Omega_m /0.3)^{0.5}$ using the joint peak and power spectrum summary statistics. If the marginal distributions for each independent density estimation were in disagreement, this would be evidence that we had insufficient forward-modelled simulations to constrain the density. The marginal distributions are generally in good agreement, albeit with some scatter.

If we were to run more simulations, we could use the scatter between the NDEs in the ensemble to make decisions about where to run new simulations, a process known as {\textit{active learning}}. Such an acquisition procedure can reduce the number of simulations necessary for robust likelihood-free inference~\citep{delfi2} and will be an important tool for large-scale implementation in current and upcoming galaxy survey analysis. 

\subsection{Posterior probabilities}

The chosen prior is uniform with respect to the physical parameters $\Omega_m$ and $\sigma_8$, but still bounded by $0.1<\Omega_m < 0.8$ and $0.45 < \sigma_8 (\Omega_m /0.3)^{0.5}< 1.05$. This prior range is the unshaded region in Figure~\ref{fig:sim_points}. To ensure this prior is used, evaluating in terms of $\Omega_m$ and  $S_8 =  \sigma_8 (\Omega_m /0.3)^{0.5}$ and applying a prior $p(S_8) \propto (\Omega_m/0.3)^{-0.5} $ gives the same posterior as evaluation of the learned likelihood in terms of $\Omega_m$ and  $\sigma_8$ directly.

As the posterior is in low dimension, evaluation on a grid (96$\times$96) is much simpler than Markov chain Monte Carlo (MCMC) sampling. The final smooth posterior distributions use {\textsc{ChainConsumer}}~\citep{Hinton2016} Kernel Density Estimation with the evaluated posterior grid points.

The left panel of Figure~\ref{fig:joint_post} shows the posterior probability for the two unknown parameters from the compressed power spectrum using the weak lensing map from DES SV data.  The right panel shows the posterior probability for the unknown parameters from the compressed peak count summary statistic using the DES SV weak lensing map. The central panel shows the parameter posterior probability distribution from the compressed joint power spectrum and peak count summary statistic. 

The two-dimensional posterior for our peak $n(\kappa)$ statistic (right panel) from our observed data is centred with low $\Omega_m$ and high $\sigma_8$, and is therefore sharply cut by the lower limit of the prior $p(\Omega_m)$. The resulting marginal posterior distribution for $\Omega_m$ is ostensibly more sharply peaked, but this is due to the prior boundary rather than the data constraints.

The parameter constraints for the combined summary statistics (central panel) are modestly improved relative to the power spectrum alone. The change in marginal posterior with respect to the parameter combination $S_8 = \sigma_8 \ (\Omega_m / 0.3)^{0.5}$ is shown in Figure~\ref{fig:comparitive}.

{
We can compare the resulting marginal in Figure~\ref{fig:comparitive} with the main result from the original shear two-point correlation DES analysis \citep{shear_des_sv}, which gave a marginal $S_8 = 0.81 \pm 0.06$ and is, therefore, completely consistent. Figure 2 of \cite{shear_des_sv} shows the two-dimensional posterior distribution, which appears in agreement to our power-spectrum result (\textit{left panel} Figure~\ref{fig:joint_post}), but cannot be directly compared as the summary statistics, including data selection effects (e.g. scale cuts) and modelling choices, are different. The joint posterior from original DES shear peak paper~\citep{des_sv_peaks} is even less directly comparable, as there are significant differences with error modelling, map making, and the definition of the peak count summary statistic, though the inferred parameters are not discrepant with our result or with \cite{shear_des_sv}.
}
\section{Results: mass map deep compressor} \label{sec:cnn_result}
\subsection{Overview}

In this section we present the likelihood-free inference results using compressed summary statistics $\mathbf{t}$ directly extracted from the DES SV weak lensing mass map using deep convolutional neural networks, the deep compressor. As described~\ref{sec:cnn_intro}, CNNs are flexible feature extractors that can be optimized as to find maximally informative summary statistics from the mass map.

We first describe our CNN architecture, the ResNet-18 model. This acts as the function that takes the noisy KS mass map as input and returns the compressed summaries. This network is included in the SSELFI implementation.

This implementation uses the MSE loss function to train the network (as described in Section~\ref{sec:compression}) and also the VMIM as an optimization objective (described below) to extract optimally informative summaries of the mass map.

As in the previous section, we use pyDELFI estimate the density $p(\mathbf{t}|\boldsymbol{\theta})$. The likelihood is then evaluated for the compressed summaries of the observed data $\mathbf{t}_o$ and combined with the prior probability distribution $p(\boldsymbol{\theta})$ to give our posterior probability for the parameters given deep compressor summary statistics extracted from the DES SV weak lensing mass map.

\subsection{Convolutional neural network architecture}

As discussed in Section~\ref{sec:cnn_intro}, we base the deep compressor method on a standard deep ResNet architecture~\citep{resnet}. Specifically, we adopt a ResNet-18 model. The main component of this architecture is the \textit{ResNet block}, in which a shortcut connection directly connects the input of the block to its output while on a parallel second branch the input is processed through several convolution layers (with associated non-linear activation functions) before being merged with the shortcut branch at the output of the block. This CNN architecture has proven extremely efficient, and enables the training of extremely deep models, with over 1000 layers.  

{ The residual structure of ResNet alleviates one of the main limitations that prevents very deep neural networks training efficiently: vanishing gradients \citep{gradient_bengio, hochreiter1997long}. Gradient back-propagation (see~\citealt{deep_learning}) is hindered by the convolutional and non-linearity layers, to the point that in a standard CNN architecture deeper than about 10 layers, the upper layers of the model (close to the input) may not receive enough gradient signal to properly train. A residual network does not face this issue as gradients always have an almost unhindered path to reach any layer of the network model through this residual connection. Another aspect that helps explain the superiority of ResNet in practice is an easier initialisation of the network parameters \citep{he2015delving}. As each individual layer of a ResNet typically has to model small residual changes between input and outputs of the residual block, a random zero-mean initialization is already an appropriate choice. 
}

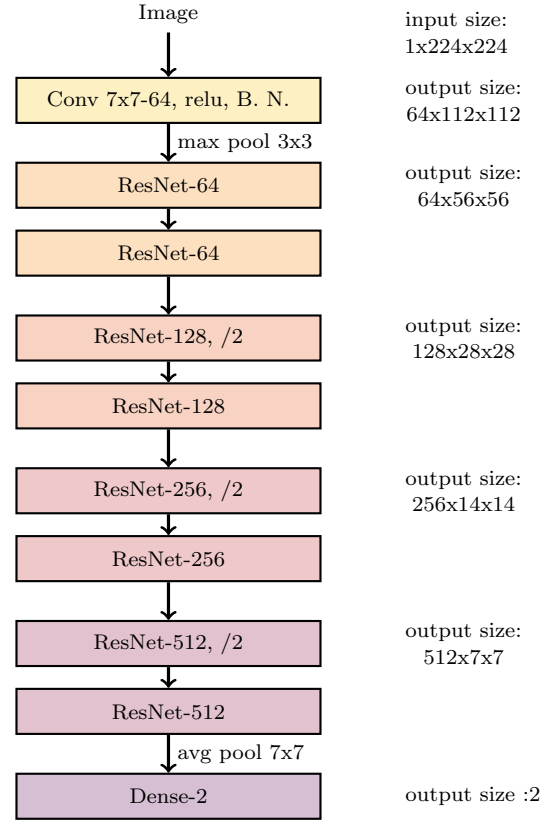
\begin{figure}
\centering
		\begin{tikzpicture}[  
	block/.style    = {draw, thick, rectangle, minimum height = 2em, minimum width = 4cm, node distance = 0.9cm},
    sum/.style      = {draw, circle, node distance = 1.cm}, 
	input/.style    = {coordinate, node distance = 1.cm}, 
  	output/.style   = {coordinate, node distance = 1.cm} 
]
\definecolor{mycolor1}{HTML}{f7d13c}
\definecolor{mycolor2}{HTML}{f8992a}
\definecolor{mycolor3}{HTML}{e96933}
\definecolor{mycolor4}{HTML}{ca484a}
\definecolor{mycolor5}{HTML}{a03360}
\definecolor{mycolor6}{HTML}{75266c}
\definecolor{mycolor7}{HTML}{4a1968}
\definecolor{mycolor8}{HTML}{1f1541}
\node[input] (input1)  {Image};
\node[block, below of=input1, fill=mycolor1!30!white]  (layer1)   {Conv 7x7-64, relu, B. N.};
\node[block, below = 0.5 of layer1, fill=mycolor2!30!white]  (layer2)   {ResNet-64};
\node[block, below of=layer2, fill=mycolor2!30!white]  (layer4)   {ResNet-64};

\node[block, below = 0.5 cm of layer4, fill=mycolor3!30!white]  (layer5)   {ResNet-128, /2};
\node[block, below of=layer5, fill=mycolor3!30!white]  (layer7)   {ResNet-128};.

\node[block, below = 0.5cm of layer7, fill=mycolor4!30!white]  (layer8)   {ResNet-256, /2};
\node[block, below of=layer8, fill=mycolor4!30!white]  (layer10)   {ResNet-256};

\node[block, below =0.5cm of layer10, fill=mycolor5!30!white]  (layer11)   {ResNet-512, /2};
\node[block, below of=layer11, fill=mycolor5!30!white]  (layer13)   {ResNet-512};

\node[block, below =0.5cm of layer13, fill=mycolor6!30!white] (layer17)  {Dense-2};

\draw[->, very thick] (input1) node [above] {Image} --  (layer1.north);
\draw[->, very thick] (layer1) -- node [right] {max pool 3x3} (layer2);
\draw[->, very thick] (layer2) -- (layer4);
\draw[->, very thick] (layer4) -- (layer5);
\draw[->, very thick] (layer5) -- (layer7);
\draw[->, very thick] (layer7) -- (layer8);
\draw[->, very thick] (layer8) -- (layer10);
\draw[->, very thick] (layer10) -- (layer11);
\draw[->, very thick] (layer11) -- (layer13);

\draw[->, very thick] (layer13) -- node [right] {avg pool 7x7} (layer17);

\node[right = 3cm of input1,align=center] {input size:\\1x224x224};
\node[right = 1cm of layer1,align=center] {output size:\\64x112x112};
\node[right = 1cm of layer2,align=center] {output size:\\64x56x56};
\node[right = 1cm of layer5,align=center] {output size:\\128x28x28};
\node[right = 1cm of layer8,align=center] {output size:\\256x14x14};
\node[right = 1cm of layer11,align=center] {output size:\\512x7x7};
\node[right = 1cm of layer17,align=center] {output size :2};

\end{tikzpicture}
	\caption{Architecture of the ResNet deep compressor. This network compresses the Kaiser-Squires weak lensing map into a two-dimensional summary statistic, which is constructed to be informative with respect to our two unknown cosmological parameters. These informative summaries are evaluated for all the many simulated mock data maps and the single observed data maps to be used for the likelihood-free inference step.}
	\label{fig:cnn_architecture}
\end{figure}

The ResNet-18 model begins with an initial single convolution layer with a larger 7$\times$7 pixel kernel with a rectified linear unit (ReLU) activation and batch normalization. The following convolutional layers have a 3$\times$3 pixel kernel size; again, each with ReLU activation and batch normalization. The input KS map is 224$\times$224, with a border region removed as a simple solution to deal with edge effects. The full network is shown in Figure~\ref{fig:cnn_architecture}.

Once the network is trained, the two-dimensional output of the model acts as our CNN compressed summary statistic. 

Our implementation is publicly available (see Appendix~\ref{append:codes}) and is based on the official TensorFlow ResNet implementation\footnote{\url{https://github.com/tensorflow/tpu}} for Google's Tensor Processing Units (TPUs). Training such a ResNet on TPU allows us to reach a high throughput of over 5500 examples per second, bringing training time to just under 4 hours for 110,000 training steps on the free Google Colab service. 

\subsection{Optimization objective}

Training the compressor to yield informative statistics can be done in several ways. In the simplest approach, we could train the network to perform regression of the cosmological parameters under either an $L_2$ norm (MSE) or $L_1$ norm loss. As explained in \ref{sec:compression}, these correspond to training the model to estimate the mean and median of the posterior distribution respectively. This corresponds to the approach followed in a number of previous works (described in Section~\ref{sec:cnn_intro}) of using CNNs for convergence maps analysis. This does not necessarily ensure the recovery of maximally informative summary statistics in general, though for fixed fiducial parameter values and assumptions of Gaussianity this may be true (Section~\ref{sec:compression}).

Another approach to train the deep compressor model is to attempt to maximise the mutual information $I(\mathbf{t}, \boldsymbol{\theta})$ between the output summary statistics and cosmological parameters. The mutual information quantifies how much knowledge about $\boldsymbol{\theta}$ is obtained by an observation $\mathbf{t}$. The mutual information can be formally defined with respect to the Kullback-Leibler divergence~\citep{kullback1951} as: 

\begin{equation} \label{eq:MI}
\begin{split}
I(\mathbf{t}, \boldsymbol{\theta})  &=  D_{\mathrm{KL}}( p(\mathbf{t}, \boldsymbol{\theta}) \parallel p(\mathbf{t}) p(\boldsymbol{\theta})) \\
        &= \int {\mathrm{d}}^n \boldsymbol{\theta} \  {\mathrm{d}}^n \mathbf{t} \ p(\mathbf{t}, \boldsymbol{\theta}) \log \left( \frac{ p (\mathbf{t}, \boldsymbol{\theta})}{p(\mathbf{t}) p(\boldsymbol{\theta})} \right) \\
		& = \int{\mathrm{d}}^n \boldsymbol{\theta} \  {\mathrm{d}}^n \mathbf{t} \  p(\mathbf{t}, \boldsymbol{\theta})  \log\left( \frac{p(\boldsymbol{\theta} | \mathbf{t})}{p(\boldsymbol{\theta})} \right) \\
		& = \int {\mathrm{d}}^n \boldsymbol{\theta} \  {\mathrm{d}}^n \mathbf{t} \ p(\mathbf{t}, \boldsymbol{\theta}) \log p(\boldsymbol{\theta} |\mathbf{t}) - \int {\mathrm{d}}^n \boldsymbol{\theta} \  p(\boldsymbol{\theta}) \log p(\boldsymbol{\theta}) \\
		& = \mathbb{E}_{p(\mathbf{t}, \boldsymbol{\theta})} \left[ \log p( \boldsymbol{\theta} | \mathbf{t}) \right] - \mathbb{E}_{p(\boldsymbol{\theta})} \left[ \log p(\boldsymbol{\theta}) \right] \\
		& =  \mathbb{E}_{p(\mathbf{t}, \boldsymbol{\theta})} \left[ \log p( \boldsymbol{\theta} | \mathbf{t}) \right] - H(\boldsymbol{\theta}) \;, 
\end{split}
\end{equation}

\noindent where $p(\mathbf{t}, \boldsymbol{\theta})$ is the joint distribution of summary statistics and cosmological parameters, which is sampled by the simulations, and the expectation value is with respect to samples $\boldsymbol{\theta}$ and  $\mathbf{t}$. On the right hand side of the second expression, we recognise the entropy $H(\boldsymbol{\theta})$ of the distribution of cosmological parameters in the set of simulations.

In the context of data compression, a compressed $\mathbf{t}$ is obtained from a realization $\mathbf{d}$ of the high-dimensional signal (in this case the full lensing mass map). In this case we parameterize this mapping as $\mathbf{t} = F_{\boldsymbol{\varphi}}(\mathbf{d})$ using our ResNet-18 model. Our goal is therefore to find the parameters $\boldsymbol{\varphi}$ of the deep compressor that maximize the mutual information between summary statistics and cosmological parameters, given by

\begin{equation}
    \boldsymbol{\varphi}^* = \argmax_{\boldsymbol{\varphi}} \  I( F_{\boldsymbol{\varphi}} (\mathbf{d}) , \boldsymbol{\theta}) \ .
\end{equation}

Unfortunately, the mutual information as expressed in equation~\ref{eq:MI} is not tractable, and estimation of this quantity remains an open problem in statistics and machine learning. However, the topic of mutual information estimation has attracted significant attention in 
the machine learning literature recently \citep[e.g.][]{Tishby2015,Chen2016,Alemi2016}, due in part to its potential for representation learning. These recent work have in common that they rely on tractable bounds on the mutual information, which allows for the mutual information to be optimized for instance in the context of training a deep neural network.

A variety of bounds exist with various properties, and we direct the interested reader to a recent review \citep{Poole2019}, but in this work we adopt the \citealt{Barber2003} variational lower bound. This is given by

\begin{equation}
    I(\mathbf{t}, \boldsymbol{\theta}) \geq \mathbb{E}_{p(\mathbf{t}, \boldsymbol{\theta})} \left[ \log q(\boldsymbol{\theta} | \mathbf{t}; \boldsymbol{\varphi}') \right] - H(\boldsymbol{\theta}) \ \ ,
\end{equation}

\noindent where $q(\boldsymbol{\theta} | \mathbf{t}; \boldsymbol{\varphi}')$ is a variational conditional distribution, which aims to model the posterior $p(\boldsymbol{\theta} | \mathbf{t})$. This lower bound becomes an equality when $q(\boldsymbol{\theta} | \mathbf{t}; \boldsymbol{\varphi}')$ matches the true posterior $p(\boldsymbol{\theta} | \mathbf{t})$. Using this bound, and taking advantage of the fact that $H(\boldsymbol{\theta})$ is constant for a given training set, we can restate the optimization problem as:

\begin{equation}
    \argmax_{\boldsymbol{\varphi},\boldsymbol{\varphi}'} \mathbb{E}_{p(\mathbf{d},\theta)} \left[ \log q \big( \boldsymbol{\theta} | F_{\boldsymbol{\varphi}} (\mathbf{d} ); {\boldsymbol{\varphi}'} \big)  \right] 
\end{equation}

This procedure is known as Variational Mutual Information Maximization (VMIM), and the optimization problem can be solved by gradient descent over the weights of the neural network $F_{\boldsymbol{\varphi}}$, and parameters of the variational distribution $q (\boldsymbol{\varphi}')$.

In practice, to train the neural compressor under VMIM, we use a conditional Normalizing Flow to model $q ( \boldsymbol{\theta} | \mathbf{t} ; {\boldsymbol{\varphi}'} )$. We adopt a 4 stage MAF, each stage containing two hidden layers of size 128, and we train jointly the concatenation of the ResNet-18 and this Normalizing Flow model under the loss:

\begin{equation}
    {J}_{\rm VMIM} (\boldsymbol{\varphi}, \boldsymbol{\varphi}') = - \sum_{n=1}^N \log q \big( \boldsymbol{\theta}^+_n | F_{\boldsymbol{\varphi}} (\mathbf{d}^+_n ); {\boldsymbol{\varphi}'} \big) \ \ ,
\end{equation}

\noindent where the sum is over the samples $\{\boldsymbol{\theta}^+_n, \ \mathbf{d}^+_n \}$ from the augmented training set.

After training, we discard the trained density estimator $q (\boldsymbol{\varphi}')$ and only export the neural compressor $F_{\boldsymbol{\varphi}}$. This is used to then compress the KS maps to summary statistics which are used in the pyDELFI likelihood-free framework described in previous sections.

\begin{figure}
\hspace*{-0.5cm}
\includegraphics[width=0.513\textwidth]{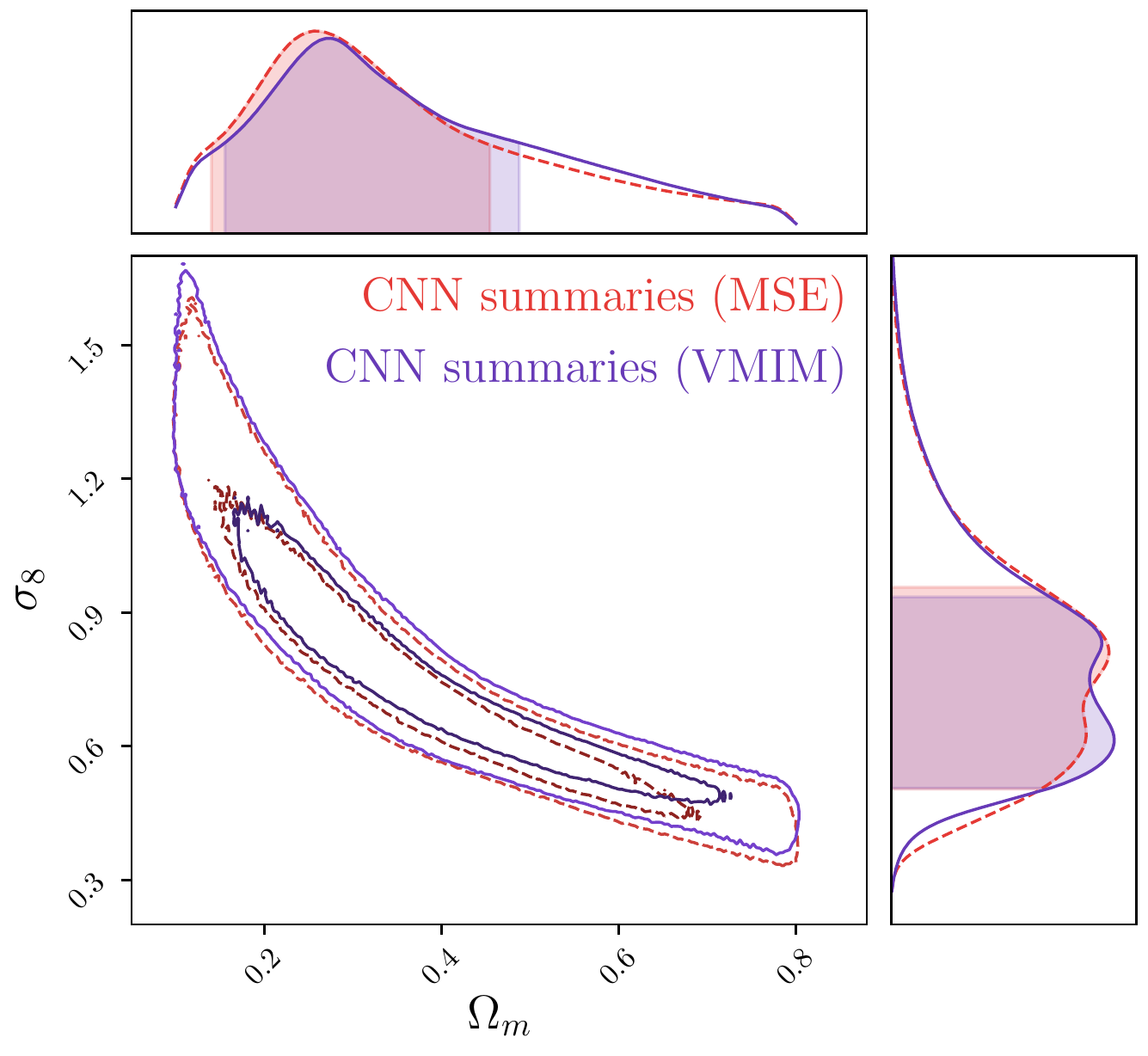}
\caption{\label{fig:neural_cnn} The posterior probability distributions from the two CNN map compressed statistics with the MSE (\textit{red dashed line}) and VMIM (\textit{purple solid line}) loss function.}
\end{figure} 

\subsection{Posterior constraints from deep CNN compression of the DES SV mass maps}

Figure~\ref{fig:neural_cnn} shows the posterior probability distributions from the DES SV weak lensing mass map using two CNN compressed map summary statistics. Each CNN had been trained using the same augmented training data but using different loss functions: the MSE and the VMIM.

As with the neural compression of the power spectrum and peak summary statistics, the network was trained and its weights fixed before application to data. Freezing the network architecture and weights before using the observed data removes the opportunity to take advantage of randomness in training, where one could, consciously or not, retrain in the hope of getting a different result due to scatter (a form of confirmation bias). This procedure of fixing the weights is a form of {\textit {blinding}}.

The CNN map compressed summary statistics (both MSE and VMIM) have slightly tighter constraints in the marginal posterior shown in Figure~\ref{fig:comparitive} than the other summary statistics. These deep compressor summaries result in a higher posterior mean (for example, comparing the joint power$\times$peaks with VMIM) for the $S_8 = \sigma_8 \ (\Omega_m / 0.3)^{0.5}$ parameter combination, but the posteriors are clearly not in tension in either the $S_8$ marginal (where the joint power$\times$peak 2-$\sigma$ credible interval contains the value of the VMIM mean posterior) or in the full two-dimensional posterior (compare Figures~\ref{fig:joint_post} and~\ref{fig:neural_cnn}).

\section{Discussion \& Conclusions}

In this work we have used likelihood-free inference to estimate the posterior probability distributions of unknown cosmological parameters given observed DES SV weak lensing map statistics. This likelihood-free framework used a forward modelling approach to include the underlying physics and the combined effects of multiple sources of data and measurement noise. 

The mock data summary statistics, labelled with their associated cosmological parameters, were compressed and compared with the observed data (also compressed) to estimate the likelihood $p( \mathbf{d_o} | \boldsymbol{\theta})$ for the given compressed summary statistic. With this simulation-based likelihood reconstruction for the DES SV data, the posterior distributions for the following summary statistics were evaluated: weak lensing map power spectrum (Figure~\ref{fig:joint_post} \textit{left panel}), weak lensing map peak count summary statistic (Figure~\ref{fig:joint_post} \textit{right panel}), the joint power$\times$peak summary statistic (Figure~\ref{fig:joint_post} \textit{centre panel}), and a deep learning compressed summary statistic of the weak lensing map using CNNs  (Figure~\ref{fig:neural_cnn}).

We use the pyDELFI~\citep{delfi2} package to perform the density estimation likelihood-free inference. To improve the efficiency of the density estimation, we aim to find some compression function taking the data summaries $\mathbf{d}$ and giving low-dimensional compressed summaries $\mathbf{t} = F(\mathbf{d})$ that retains the information about the unknown parameters. We aimed to learn such a compression function using deep neural networks (Section~\ref{sec:compression}). Compressing the weak lensing map directly, rather than compressing summaries (e.g. power spectra), is an extension of this, which was implemented using deep convolutional neural networks (Section~\ref{sec:cnn_result}).

In this work we have implemented a series of validation steps in our likelihood-free inference pipeline. Some of these tests aim to validate the forward model, which ensures the reliability of the physics modelling and the data modelling. We also validate the density estimation step, which ensures the reliability of the resulting posterior distributions given the simulated data.

To validate the forward model we compare the measured summary statistics from our simulations with summary statistics measured from higher-resolution simulations. In our case, we show that, up to the noise level of our data, the {\textsc{L-Picola}} simulations gives power spectra and peak count summary statistics that are consistent with the MICE simulations. In this framework, we can deal with {\textit{known unknowns}}. For example, we are aware that approximations in the {\textsc{L-Picola}} algorithm impact the small scales in the matter density, and therefore satisfied that the discrepancy in the power spectrum at high $\ell$ exists and is below the noise-level.

\begin{figure}
\hspace{-0.5cm}
\includegraphics[width=0.54\textwidth]{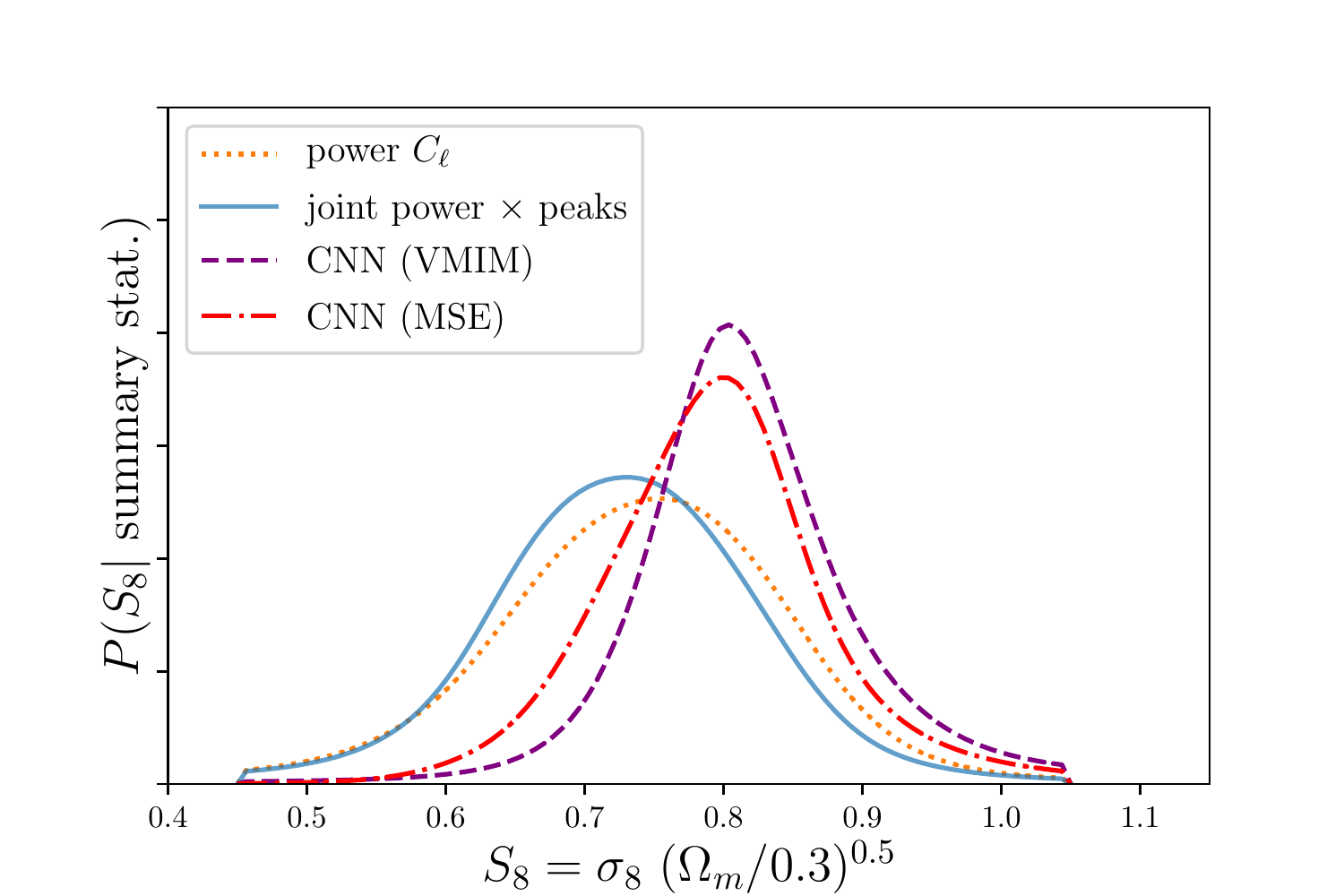}
\caption{\label{fig:comparitive} The marginal posterior probability densities for the parameter combination $S_8 = \sigma_8 \ (\Omega_m / 0.3)^{0.5}$ from the power spectrum compressed summary (\textit{dotted orange line}), the joint power spectrum \& peak compressed summary statistic (\textit{solid blue line}), and from the two CNN map compressed summary statistics with MSE loss (\textit{dashed purple line}) and VMIM loss (\textit{dash-dot red line}).}
\end{figure} 

There may be {\textit{unknown unknowns}} in the forward model that are more difficult to validate. For example, one could construct a generator that created mock data with the power spectra matching theory perfectly, but still had a theoretically incorrect power spectrum covariance. Here we must rely on the physics encoded in the simulator to reflect the true behaviour of our system (the evolution of the cosmic matter density), and test for inaccuracies that could be reasonably expected  (e.g. finite-resolution, $N$-body approximations, baryonic effects). In all parameter or model inference, one should be able to describe the data model, and the same concerns about {\textit{unknown unknowns}} should apply.

The clarity afforded by the ability to forward model mock data is a strength of likelihood-free inference. In the likelihood-free framework the individual elements of the forward model are distinct and testable, so are less easily hidden.

In this work, our forward model includes or implicitly accounts for: non-Gaussian shape noise, the non-Gaussian density field, all mask and spatially-varying shape noise effects, resulting higher-order moments of the sampling distribution, projection and flat-sky/Limber effects.

Effects that are often included as nuisance parameters, but which we have omitted, are the intrinsic alignments and the uncertainty in the photometric redshift distribution of the source galaxies. With the addition of extra nuisance parameters, we could have used different models to include the intrinsic alignments of galaxies. This is something that can be done in future work. \cite{des_sv_peaks} found that parameter constraints from peak counts can indeed be shifted by the effects of intrinsic alignments.

We could additionally include uncertainties in the photometric redshift distribution, either by simply marginalising over nuisance parameters that re-calibrate or transform the distribution (e.g.~\citealt{shear_des_sv}) at the expense of additional parameters or, more generally, including the redshift analysis in the forward model.

To validate the density estimation step, we trained an ensemble of neural density estimators with different architectures. Figure~\ref{fig:marginals} shows marginal posterior distributions for the different neural density estimators (from the compressed joint power$\times$peak data). The results from the different density estimators are in general agreement, though additional simulations with certain parameter combinations would decrease the variance of the ensemble. The model averaged stack of NDEs that is used for deriving the final posterior is more robust than any individual NDE in the ensemble.

By running additional simulations ``on the fly'', one could update the density ensemble after each new batch of simulations to learn where in parameter space to best run the next batch. {\textit{Active learning}} has been shown to be far more efficient in terms of the number of simulations needed (\citealt{bolfi};~\citealt{delfi2}). For likelihood-free inference using $N$-body simulations from data from current and upcoming surveys, with much larger volumes and a larger parameter space, this approach may be vital.

The likelihood-free approach can not only account for non-Gaussianity (which is not just a problem for the higher-order statistics in weak lensing), but it also  provides an efficient and alternative analysis pipeline that avoids many of the troublesome aspects of the standard cosmological inference pipelines. For example, with the same number of simulations that may be used to estimate the inverse covariance matrix for a Gaussian likelihood, we have estimated the full sampling distribution of the compressed summaries.

The likelihood-free framework improves the reliability of inference from standard summary statistics (e.g. power spectra) by providing a flexible and robust way forward by using direct comparison of simulations with data. It can also allow us, without resorting to misleading likelihood approximations, to use non-standard summary statistics of our data (including peaks and deep CNN map statistics) that can extract information beyond the standard summaries. { As well as opening up non-standard summary statistics, likelihood-free inference for weak lensing surveys may also enable us to extract information from non-standard weak lensing observables (e.g. magnification, \citealp{Hildebrandt2009,VanWaerbeke2010,Hildebrandt2013,Duncan:2013bb,heavens2013combining, Alsing2015}) that have been inhibited by complicated systematics effects that could (in principle) be included in forward simulations. }

\section*{Acknowledgements}

The authors thank Lorne Whiteway and Pablo Lemos for helpful comments on the manuscript, and thank Ofer Lahav and Benjamin Wandelt for useful discussions. NJ acknowledges funding from the {\'E}cole Normale Sup{\'e}rieure (ENS) and also acknowledges support from STFC Grant ST/R000476/1.

Some of the results in this paper have been derived using the \texttt{healpy} package~\citep{Zonca2019}. We also acknowledge use of \texttt{matplotlib}~\citep{matplotlib}, \texttt{keras}~\citep{keras}, \texttt{TensorFlow}~\citep{tf}, and \texttt{chainconsumer}~\citep{Hinton2016}.

\section*{Data availability}

The data underlying this article are publicly available from the \texttt{DES Data Management} system as part of the SVA1 Gold Release: \url{https://des.ncsa.illinois.edu/releases/sva1}

{
We have made the simulations used, along with associated code, publicly available: \github~  \url{https://github.com/NiallJeffrey/Likelihood-free_DES_SV}. Links to packages used in this paper are including in Appendix~\ref{append:codes}.}




\bibliographystyle{mnras}
\bibliography{des_lfi}



\appendix

\section{Summary statistic neural compression training} \label{append:neural}

For the neural compression of the power and peak summary statistics, we tested a series of architectures with varying activation functions, loss functions, and learning rates. The final (best performing) architecture used three dense hidden layers with 100 nodes, each followed by a ReLU activation function, and a final dense layer (to a two element output) without an activation function. The network was trained using the stochastic optimizer \textit{Adam}~\citep{kingma2014adam} with 130000 training samples and 55000 validation samples (generated using the data augmentation method described in Section~\ref{sec:compression}). The network, which showed no evidence of significant overfitting, was trained for 20 epochs.

For the loss function we considered two options: the $L_1$ loss and the $L_2$ MSE loss (e.g. equation~\ref{eq:mse_loss}). As discussed in Section~\ref{sec:compression}, the $L_2$ minimization corresponds to a point estimate of the posterior mean and the $L_1$ minimization corresponds to a point estimate of the posterior mode. From the augmented training data, a small sample of four realizations were taken as mock observations and each compressed in two different ways (with $L_1$ and $L_2$ trained networks). The compressed summaries were used in our pyDELFI pipeline with the result that the $L_2$ MSE loss consistently gave slightly tighter constraints (with the four mock observations) than the $L_1$ loss. Any differences between the two choices of loss function were nevertheless extremely small.

All of these tests were performed before the final likelihood-free inference step using data, to avoid misleading results due to post-hoc analysis (analogous to ``p-hacking'').

\section{Public code} \label{append:codes}

In this work we have used code to generate mock simulations, compress the observed summary statistics of the data, and estimate the likelihood-free posterior probability densities.

\begin{itemize}
    \item To generate the mock simulations we use the {\textsc{L-Picola}} dark matter simulation code~\citep{picola}: {\url{https://cullanhowlett.github.io/l-picola/}}.

    \item To convert the dark matter overdensity shells into convergence maps we use a Born approximation ray tracing code, which we have made available: {\url{https://github.com/NiallJeffrey/born_raytrace}}.

    \item For density estimation likelihood-free inference we use the pyDELFI code~\citep{delfi2}: {\url{https://github.com/justinalsing/pydelfi}}.

    \item For the deep compressor CNN summary statistic extraction we use SSELFI: {\url{https://github.com/EiffL/SSELFI}}.

\end{itemize}


\bsp	
\label{lastpage}
\end{document}